\documentclass[12pt]{article}

\usepackage{amsfonts}
\usepackage{amsmath}
\usepackage{esint}
\usepackage[margin=1in]{geometry}
\usepackage{graphicx}
\usepackage{epstopdf}
\usepackage{amssymb}
\usepackage{cite}
\usepackage{multicol}
\usepackage{color}
\usepackage[section]{placeins}
\usepackage[linesnumbered,ruled,vlined]{algorithm2e}
\usepackage{framed}
\usepackage[font=small,labelfont=bf]{caption}
\usepackage{hyperref}

\newcommand{\ds} {\displaystyle}
\newcommand{\Frac}[2]{\ds \frac{#1}{#2}}
\newcommand{\R}{{\mathbb R}}

\title{When Artificial Parameter Evolution Gets Real: Particle Filtering for Time-Varying Parameter Estimation in Deterministic Dynamical Systems}
\author{Andrea Arnold*}
\date{}

\begin{document}
\maketitle

\small 
\centerline{Department of Mathematical Sciences, Worcester Polytechnic Institute, Worcester, MA, USA}
\vspace{.2cm}

\centerline{*correspondence: anarnold@wpi.edu}

\normalsize

\bigskip

\begin{abstract}

Estimating and quantifying uncertainty in unknown system parameters from limited data remains a challenging inverse problem in a variety of real-world applications.
While many approaches focus on estimating constant parameters, a subset of these problems includes time-varying parameters with unknown evolution models that often cannot be directly observed.
This work develops a systematic particle filtering approach that reframes the idea behind artificial parameter evolution to estimate time-varying parameters in nonstationary inverse problems arising from deterministic dynamical systems.  
Focusing on systems modeled by ordinary differential equations, we present two particle filter algorithms for time-varying parameter estimation: one that relies on a fixed value for the noise variance of a parameter random walk; another that employs online estimation of the parameter evolution noise variance along with the time-varying parameter of interest. 
Several computed examples demonstrate the capability of the proposed algorithms in estimating time-varying parameters with different underlying functional forms and different relationships with the system states (i.e., additive vs. multiplicative).\\

\end{abstract}



\section{Introduction}

Many applications in modern day science and engineering involve system parameters that are unknown and must be estimated using little to no prior information.
In particular, when using mechanistic models to analyze and predict the behavior of these systems, the inverse problem of estimating and quantifying uncertainty in the unknown model parameters from limited data remains a big challenge.

While many approaches focus on estimating constant (or static) parameters, a subset of these problems includes parameters that are known to vary with time but have unknown dynamics that often cannot be directly observed. 
Examples include the seasonal transmission parameter in modeling the spread of infectious diseases \cite{Altizer2006, Grassly2006}, the input stimuli in modeling neuron dynamics \cite{Campbell2020, Voss2004}, and the tissue optical properties in modeling thermal laser-tissue interactions \cite{ArnoldFichera2022}.  
The main challenge in estimating time-varying parameters lies in accurately accounting for their time evolution without information regarding their temporal dynamics.  
In this sense, time-varying parameters can be thought of as unmeasured system states with unknown evolution models.

The aim of this work is to develop a systematic approach and computational framework for estimating time-varying parameters in nonstationary inverse problems arising from deterministic dynamical systems.  
More specifically, we consider systems modeled by ordinary differential equations (ODEs) of the form
\begin{equation}\label{Eq:ODE_model}
 \Frac{dx}{dt} = f(t,x,\theta), \qquad x(0) = x_0,
\end{equation}
where $x = x(t) \in\R^d$ represents the vector of model states, $\theta=\theta(t)\in\R^p$ is the vector of unknown time-varying parameters, $f:\R\times\R^d\times\R^p\to\R^d$ is a known model function describing the state dynamics, and $x_0$ is the (possibly unknown or poorly known) vector of the initial states.
While the model in \eqref{Eq:ODE_model} might also depend on constant parameters, we emphasize here the dependence of the system on unknown time-varying parameters and assume that any constant parameters are known and fixed.

Further assume that a sequence of discrete, noisy observations $y_j\in\R^m$, $j=1, 2, \dots$, is available, relating back to the system states via the model
\begin{equation}\label{Eq:Obs_model}
 y_j = g(x(t_j)) + w_j, \qquad 0<t_1<t_2<\cdots
\end{equation}
where $g:\R^d\to\R^m$, $m\leq d$, is a known function and $w_j$ represents the observation error.  
While $g$ could generally be nonlinear, in this work we assume that the state vector $x(t)$ is at least partially observable and that $g$ is a linear mapping distinguishing between the observable and unobservable components.
Given the observed data, our goal is to estimate the time-varying parameter vector $\theta(t)$ along with the state vector $x(t)$ at some discrete times.

Bayesian filtering methods provide a natural framework for tackling dynamic inverse problems of this type, given the sequential nature of the observations and time-dependent quantities of interest.
In this work, we focus specifically on particle filters, also referred to as sequential Monte Carlo methods, which use importance sampling within sequential updating schemes to evolve samples from the posterior distribution $\pi(x,\theta | y)$ of the model states and parameters conditioned on the available observations.
The updates are performed by defining a statistical evolution-observation model that corresponds with the deterministic set-up in \eqref{Eq:ODE_model}--\eqref{Eq:Obs_model}, using discrete-time stochastic equations to describe the system states and observations. 
Particle filters have been widely employed for static parameter estimation in state-space models (see, e.g., \cite{Kantas2009, Kantas2015} and the references therein for an overview), with some methods specifically designed for application to systems of differential equations \cite{Arnold2013, Arnold_thesis}.
Particle filtering with differential equations requires special attention when propagating the particles forward in time, in particular with respect to the numerical solution to the ODE system \eqref{Eq:ODE_model} and related computational considerations.

In most particle filtering approaches, the goal is to design a sampling scheme to perform the sequential posterior update
\[
\pi(x_j, \theta | y_j) \ \longrightarrow \ \pi(x_{j+1}, \theta | y_{j+1})
\]
where the parameter vector $\theta$ is typically assumed to be constant in time: 
If $\theta$ is known, the parameters are fixed to their specified values and the dependence on $\theta$ is dropped; the filtering problem that remains is referred to as the state estimation problem.
If $\theta$ is unknown, the parameters can be estimated by applying artificial time evolution, i.e., perturbing the parameter values from time $j$ to $j+1$ using a random walk model, with the goal of finding a joint distribution informing their most likely constant values given the available data.
This static parameter evolution is considered to be ``artificial'' since the parameters are constant and therefore, in the sense of differential equations, $d\theta/dt = 0$.

We consider instead unknown parameters that actually do evolve with time, i.e., $\theta = \theta(t)$ with $d\theta/dt=h(t)$ for some unknown mapping $h:\R\to\R^p$, and aim to perform the posterior update
\[
\pi(x_j, \theta_j | y_j) \ \longrightarrow \ \pi(x_{j+1}, \theta_{j+1} | y_{j+1})
\]
where our goal is to estimate $\theta(t)$ along with the system states.
If $d\theta/dt = h(t)$ was known, this problem could be addressed by treating $\theta(t)$ as an additional state of the ODE model in \eqref{Eq:ODE_model} and applying a particle filtering technique for state estimation.
However, since we do not know the mapping $h$ and thereby do not have an expression for $d\theta/dt$, we must determine a surrogate model for the parameter time evolution.

One approach is to assign an approximation model for $\theta(t)$, e.g., using piecewise constant or spline functions, and then applying static parameter estimation techniques to estimate the unknown approximation model coefficients. 
This approach has been applied successfully in previous work modeling periodic time-varying parameters \cite{Arnold2018, Arnold2020}; however, it enforces a functional form for $\theta(t)$ that may not accurately reflect its true dynamics, especially in online estimation cases where the form of $\theta(t)$ may change drastically over time.
To this end, we propose a more flexible parameter tracking framework that leverages particle filtering methodology and extends the idea behind artificial parameter evolution to actually evolve time-varying parameters.

In this work, we develop a particle filtering approach for time-varying parameter estimation that employs random walk modeling to track the underlying true evolution of the parameters.
We focus on time-varying parameters in differential equations models, where the nonlinear state evolution requires computing the numerical solution to the system \eqref{Eq:ODE_model} at each time step, and highlight the considerations needed for practical implementation.
As we will demonstrate in the numerical results, the proposed algorithms allow for the systematic tracking of parameters with a variety of functional forms, where the particle filter sample mean is used as an estimate for the parameter at each time and the corresponding credible intervals provide a natural measure of uncertainty.

Previous work in \cite{Nemeth2014} proposes a sequential Monte Carlo method to estimate rapidly changing, piecewise constant parameters in target tracking applications, where the parameters change abruptly at an unknown set of time instances and remain constant in between.
The work in \cite{Cheng2017} uses a kernel density-based particle filter to estimate multipath biases in an application to global navigation satellite systems, provided a known linear operator for state evolution.
The work in \cite{Stano2014} compares the application of a bootstrap particle filter and feedback particle filter in estimating a time-varying soil-dependent parameter in a stochastic model for sedimentation.
In the setting of ODEs, several recent works have addressed the time-varying parameter estimation problem using variants of the augmented ensemble Kalman filter \cite{Campbell2020, Arnold2019}.
These works employ random walk models with fixed noise variances to account for the parameter evolution. 
Results emphasize the importance of choosing an appropriate value for the random walk noise variance in order to well track the parameter and avoid filter divergence.

The work in this paper contributes two particle filtering algorithms for time-varying parameter estimation in deterministic dynamical systems: the first relies on a fixed value for the noise variance of the parameter random walk; the second accommodates online estimation of the noise variance along with the time-varying parameter of interest. 
Numerical results in several computed examples demonstrate the algorithms' capability in estimating time-varying parameters with a variety of underlying functional forms (e.g., slowly vs. rapidly varying, continuous vs. discontinuous with jump discontinuities) and different influences on the system (i.e., additive vs. multiplicative relationships with the model states).

The remainder of the paper is organized as follows:
Section~\ref{Sec:PF_State} reviews particle filtering for state estimation, highlighting the aspects specific to nonstationary inverse problems with differential equations models.
Section~\ref{Sec:PF_TVP} details the proposed particle filtering algorithm for time-varying parameter estimation and corresponding implementation considerations, along with a computed example.
Section~\ref{Sec:PF_TVP_Enhanced} describes an enhanced algorithm modified to include simultaneous online estimation of the parameter evolution noise variance and several computed examples.
Section~\ref{Sec:Discussion} gives a discussion of the methods and results,
and Section~\ref{Sec:Conclusions} concludes the manuscript.



\section{Particle Filtering For State Estimation}
\label{Sec:PF_State}

Before addressing time-varying parameter estimation, we review the main ideas behind particle filtering for state estimation and highlight the aspects needed to apply these techniques in the setting of deterministic dynamical systems.  
For an overview of particle filtering and sequential Monte Carlo methods, we refer to a number of relevant texts and review articles on the subject; see, e.g., \cite{Fearnhead2018, Li2015, Doucet2011, Godsill2019, Arulampalam2002, Cappe2007, Chopin_book, KaipioSomersalo}.
In this work, we follow similarly to the importance sampling scheme in \cite{LiuWest2001}, which utilizes the auxiliary particle technique in \cite{Pitt1999}.

We begin by assuming a continuous-time ODE model of the form in \eqref{Eq:ODE_model} that can be discretized and solved numerically via a time-marching scheme (e.g., Runge-Kutta or linear multistep methods).
Further, we assume sequential observations relating to some subset of the system states as in \eqref{Eq:Obs_model}. 
For the state estimation problem, we assume that the model parameters are known and fixed; we thereby temporarily suppress the dependence of the model and subsequent probability distributions on $\theta$.

To formulate our statistical model, we define the stochastic processes $\{X_j\}_{j=0}^\infty$ and $\{Y_j\}_{j=1}^\infty$, where the random variable $X_j\in\R^d$ represents the model states and the random variable $Y_j\in\R^m$ represents the observations at time $j$.
Following the assumptions that
\begin{itemize} 
\item[(i)] $\{X_j\}_{j=0}^\infty$ is a Markov process; i.e.,
\[
\pi(x_{j+1} | x_0,\dots,x_j) = \pi(x_{j+1} | x_j) 
\]
\item[(ii)] $\{Y_j\}_{j=1}^\infty$ is a Markov process with respect to the states; i.e.,
\[
\pi(y_{j} | x_0,\dots,x_j) = \pi(y_j | x_j) 
\] 
\item[(iii)] $\{X_j\}_{j=0}^\infty$ depends on the past observations only by way of its own history; i.e.,
\[
\pi(x_{j+1} | x_j,y_0\dots,y_j) = \pi(x_{j+1} | x_j) 
\]
\end{itemize}
we can write an evolution-observation model of the form
\begin{eqnarray}
X_{j+1} &=&  F(X_j) + V_{j+1}, \quad V_{j+1} \sim \mathcal{N}(0, \mathsf{C}_{j+1}) \label{Eq:State_Evolution} \\[0.2cm]
Y_{j+1} &=& G(X_{j+1}) + W_{j+1}, \quad W_{j+1} \sim \mathcal{N}(0, \mathsf{D}_{j+1}) \label{Eq:Observation}
\end{eqnarray}
where the state evolution equation \eqref{Eq:State_Evolution} defines the transition density $\pi(x_{j+1}|x_j)$ and the observation equation \eqref{Eq:Observation} informs the likelihood function $\pi(y_{j+1}|x_{j+1})$.  
In the setting of deterministic dynamical systems, the operator $F$ in \eqref{Eq:State_Evolution} represents the numerical solution to the ODE system in \eqref{Eq:ODE_model} from time $j$ to $j+1$.  
The operator $G$ in \eqref{Eq:Observation} defines a mapping between the model states and the observed data, which is linear in the case of direct observations of the state vector and is analogous to the function in \eqref{Eq:Obs_model}.
The state and observation noise processes, $\{V_j\}_{j=1}^\infty$ and $\{W_j\}_{j=1}^\infty$, respectively, are assumed to be mutually independent, here modeled using Gaussian random variables with zero mean and prescribed covariance matrices, $\mathsf{C}_{j+1}$ and $\mathsf{D}_{j+1}$.
While these covariance matrices may generally vary in time, we assume that the noise processes have time-invariant covariance matrices, such that $\mathsf{C}_{j+1} = \mathsf{C}$ and $\mathsf{D}_{j+1} = \mathsf{D}$.

Let $D_j = \{ y_1,y_2,\dots,y_j \}$ denote the set of accumulated observations up to time $j$.
Since observations may occur only at a subset of the discrete time instances, we let $y_j = \emptyset$ if there is no observation at time $j$.  
Starting at time $j=0$, our goal in particle filtering is to design a sequential updating scheme that allows us to track the time evolution of representative samples from the sequence of posterior distributions
\[
\pi(x_j | D_j) \ \longrightarrow \ \pi(x_{j+1} | D_{j+1})
\]
where $\pi(x_0 | D_0)$, $D_0 = \emptyset$, is a prior density encoding any known information on the initial states.
In other words, starting with an initial sample of states $\{ x_0^{(1)}, x_0^{(2)}, \dots, x_0^{(N)}\}$ and corresponding weights $\{ w_0^{(1)}, w_0^{(2)}, \dots, w_0^{(N)}\}$ that represent a Monte Carlo approximation to the prior density $\pi(x_0 | D_0)$, we aim to compute the updates
\[
\mathcal{S}_j \ \longrightarrow \ \mathcal{S}_{j+1}, \quad \text{where} \quad \mathcal{S}_j = \big\{ (x_j^{(n)}, w_j^{(n)})\big\}_{n=1}^N, \quad j = 0, 1, \dots.
\]

We perform the evolution from time $j$ to $j+1$ in two phases, a prediction step and an observation update,
\[
\pi(x_j | D_j) \ \longrightarrow \ \pi(x_{j+1} | D_j) \ \longrightarrow \ \pi(x_{j+1} | D_{j+1})
\]
by applying Bayes' theorem,
\begin{equation}\label{Eq:Bayes_Thm}
\pi(x_{j+1} | D_{j+1}) \propto \pi(y_{j+1} | x_{j+1}) \pi(x_{j+1} | D_j)
\end{equation}
where $\pi(x_{j+1} | D_j)$ is the prior density on the states at time $j+1$ and $\pi(y_{j+1} | x_{j+1})$ is the likelihood function.
From the Chapman-Kolmogorov equation, the prior density is given by 
\[
\pi(x_{j+1} | D_j) = \ds\int \pi(x_{j+1} | x_j ) \pi(x_j | D_j) dx_j
\]
and we can write its Monte Carlo approximation as
\[
\pi(x_{j+1} | D_j) \approx \ds\sum_{n=1}^N w_j^{(n)} \pi\big(x_{j+1} | x_j^{(n)}\big).
\]
Using this approximation in Bayes' theorem \eqref{Eq:Bayes_Thm} yields the formula
\begin{equation}\label{Eq:PF_post}
\pi(x_{j+1} | D_{j+1}) \propto \pi(y_{j+1} | x_{j+1}) \ds\sum_{n=1}^N w_j^{(n)} \pi\big(x_{j+1} | x_j^{(n)}\big) .
\end{equation}

Various techniques can be applied to obtain a Monte Carlo approximation of the posterior density in \eqref{Eq:PF_post}.
To utilize auxiliary particles in the style of \cite{Pitt1999}, we choose predictors $\widehat{x}_{j+1}^{(n)}$, $n=1,\dots,N$, that serve as estimates of $x_{j+1}$.
In this work, we use the forward propagation of each state particle $x_j^{(n)}$ from time $j$ to $j+1$ via the numerical solution to \eqref{Eq:ODE_model} as its predictor: 
\[
\widehat{x}_{j+1}^{(n)} = F\big(x_j^{(n)}\big), \quad n = 1, \dots, N.
\]
We evaluate the fitness of each predictor by computing the weights
\[
g_{j+1}^{(n)} = w_j^{(n)}  \pi(y_{j+1} | \widehat{x}_{j+1}^{(n)}) 
\]
which indicate how well the predictors agree with the observation $y_{j+1}$, effectively taking the likelihood into account before resampling.
The fitness weights can be incorporated into the formula in \eqref{Eq:PF_post} by moving the likelihood inside the summation and writing
\[
\pi(x_{j+1} | D_{j+1}) \propto  \ds\sum_{n=1}^N g_{j+1}^{(n)} \Frac{ \pi(y_{j+1} | x_{j+1}) }{\pi(y_{j+1} | \widehat{x}_{j+1}^{(n)}) } \pi\big(x_{j+1} | x_j^{(n)}\big) .
\]
We use the fitness probabilities to draw with replacement auxiliary indices, $\ell_n \in \{1, 2, \dots, N\}$, and reshuffle our current state particles and predictors based on these indices; i.e.,
\[
x_j^{(n)} \ \leftarrow \ x_j^{(\ell_n)} , \quad \widehat{x}_j^{(n)} \ \leftarrow \ \widehat{x}_j^{(\ell_n)} , \quad n = 1, \dots, N.
\]
We then repropagate and innovate the reshuffled particles to generate a new sample of states at time $j+1$, such that
\[
x_{j+1}^{(n)} = F\big(x_j^{(n)}\big) + v_{j+1}^{(n)}, \quad v_{j+1}^{(n)}\sim\mathcal{N}(0,\mathsf{C}), \quad n = 1, \dots, N
\]
with corresponding weights 
\[
w_{j+1}^{(n)} = \Frac{ \pi(y_{j+1} | x_{j+1}^{(n)} ) }{\pi(y_{j+1} | \widehat{x}_{j+1}^{(n)}) }.
\]

The steps of the particle filtering algorithm for state estimation are summarized below.
In particular, note that computing the state predictors in Step~1 and repropagating the state particles in Step~5 requires the numerical solution to the ODE system in \eqref{Eq:ODE_model} from time $j$ to $j+1$ for each particle, $n=1,\dots,N$.
We will discuss important implementation considerations relating to this in the next section.

\vspace{0.5cm}

\noindent\makebox[\linewidth]{\rule{\textwidth}{1pt}}\\[0.2cm]
\textbf{Algorithm:} Particle Filter for State Estimation\\[0.2cm]
Given the prior density $\pi(x_0|D_0)$, draw an initial sample of state particles with equal weights:
\[
\mathcal{S}_0 = \big\{ \big(x_0^{(n)}, w_0^{(n)} \big)\big\}_{n=1}^N, \qquad w_0^{(1)} = \cdots = w_0^{(N)} = \frac{1}{N}
\]
Set $j=0$. While $j<T$, where $T$ is the number of available observations, do:
\begin{enumerate}
\item Propagate forward to compute the state predictors: 
\[
\widehat{x}_{j+1}^{(n)} = F\big(x_j^{(n)}\big), \qquad n = 1, \dots, N
\]
\item Calculate the fitness weights for each $n$:
\[
g_{j+1}^{(n)} = w_j^{(n)}  \pi(y_{j+1} | \widehat{x}_{j+1}^{(n)}) , \qquad g_{j+1}^{(n)} \ \leftarrow \ \frac{g_{j+1}^{(n)} }{\sum_n g_{j+1}^{(n)}} 
\]
\item Draw with replacement the auxiliary indices $\ell_n \in \{1, 2, \dots, N\}$, $n=1,\dots, N$, using the fitness probabilities:
\[
P\{\ell_n = k \} = g_{j+1}^{(k)}
\]
\item Reshuffle current samples of states and predictors:
\[
x_j^{(n)} \ \leftarrow \ x_j^{(\ell_n)} , \qquad \widehat{x}_j^{(n)} \ \leftarrow \ \widehat{x}_j^{(\ell_n)} , \qquad n = 1, \dots, N
\]
\item Repropagate and innovate the reshuffled state particles:
\[
x_{j+1}^{(n)} = F\big(x_j^{(n)}\big) + v_{j+1}^{(n)}, \qquad v_{j+1}^{(n)}\sim\mathcal{N}(0,\mathsf{C}), \qquad n = 1, \dots, N
\]
\item Recalculate the weights for each $n$:
\[
w_{j+1}^{(n)} = \Frac{ \pi(y_{j+1} | x_{j+1}^{(n)} ) }{\pi(y_{j+1} | \widehat{x}_{j+1}^{(n)}) }, \qquad w_{j+1}^{(n)} \ \leftarrow \ \frac{w_{j+1}^{(n)} }{\sum_n w_{j+1}^{(n)}} 
\]
\item Set $j \leftarrow j+1$ and repeat from Step 1.
\end{enumerate}
\noindent\makebox[\linewidth]{\rule{\textwidth}{1pt}}\\

\vspace{0.5cm}



\section{From Artificial to Actual Parameter Evolution:\\ Enter Time-Varying Parameters}
\label{Sec:PF_TVP}

The previous section details a particle filtering algorithm for state estimation, under the assumption that model parameters are known and fixed.  To address time-varying parameter estimation, we now reintroduce the dependence of the model and corresponding probability distributions on unknown $\theta$.
In the context of estimating static parameters, previous works have introduced the notion of ``artificial parameter evolution'': the idea is to artificially evolve samples representing the unknown static parameters over time by adding small perturbations to their constant values in order to find distributions of their most likely values conditioned on the observed data \cite{LiuWest2001}.
Since the parameters are actually static, without additional adjustments, this approach can result in the loss of information over time such that the posterior distributions are far from the true underlying distributions with fixed parameters.

However, our goal in this work is to track the actual time evolution of time-varying parameters, $\theta=\theta(t)$.
By applying the parameter perturbation model as a proxy for the unknown parameter dynamics, we design a particle filter updating scheme to track the evolution of
\[
\pi(x_j, \theta_j | D_j) \ \longrightarrow \ \pi(x_{j+1}, \theta_{j+1} | D_{j+1})
\]  
thereby estimating the time-varying parameters along with the system states.
Here, $\theta_j$ denotes an approximation to $\theta(t_j)$, where $\theta(t)$ is the true (and unknown) parameter function, and $\pi(x_0, \theta_0 | D_0)$, $D_0 = \emptyset$, is a joint prior density encoding any known information on the initial states and time-varying parameters.

We now detail the modifications to the particle filter in Section~\ref{Sec:PF_State} needed for joint state and time-varying parameter estimation.
Starting with a joint initial sample of states and parameters $\{ (x_0^{(n)},\theta_0^{(n)})\}_{n=1}^N$ and corresponding weights $\{ w_0^{(n)}\}_{n=1}^N$ that represent a Monte Carlo approximation to the prior density $\pi(x_0,\theta_0 | D_0)$, we aim to compute the sequential updates
\[
\mathcal{S}_j \ \longrightarrow \ \mathcal{S}_{j+1}, \quad \text{where} \quad \mathcal{S}_j = \big\{ (x_j^{(n)}, \theta_j^{(n)}, w_j^{(n)})\big\}_{n=1}^N, \quad j = 0, 1, \dots
\]
via the associated two-phase scheme
\[
\pi(x_j, \theta_j | D_j) \ \longrightarrow \ \pi(x_{j+1}, \theta_{j+1} | D_j) \ \longrightarrow \ \pi(x_{j+1}, \theta_{j+1} | D_{j+1}).
\]
Applying Bayes' theorem gives
\[
\pi(x_{j+1},\theta_{j+1} | D_{j+1}) \propto \pi(y_{j+1} | x_{j+1}, \theta_{j+1}) \pi(x_{j+1}, \theta_{j+1} | D_j)
\]
where $\pi(x_{j+1}, \theta_{j+1} | D_j)$ is the joint prior density on the states and time-varying parameters at time $j+1$ and $\pi(y_{j+1} | x_{j+1}, \theta_{j+1})$ is the likelihood function.

While we have a model for state propagation via the dynamics in \eqref{Eq:ODE_model} and corresponding state evolution equation in \eqref{Eq:State_Evolution}, we do not have a known form of $d\theta/dt$ to inform the parameter evolution.  
Instead, we define a random walk perturbation model of the form
\begin{equation}\label{Eq:Param_model}
\theta_{j+1} = \theta_j + \xi_{j+1}, \quad \xi_{j+1} \sim \mathcal{N}(0, \mathsf{E}_{j+1})
\end{equation}
to serve as a surrogate parameter evolution model, where $\mathsf{E}_{j+1}$ is a specified covariance matrix relating to the parameter evolution noise process, and $\theta_j$ and $\xi_{j+1}$ are conditionally independent given $D_j$.
The noise term $\xi_{j+1}$ is sometimes referred to as the ``drift'' term in the parameter random walk, interpreting the model such that the parameter values at time $j+1$ drift away from their previous values at time $j$.
The covariance matrix $\mathsf{E}_{j+1}$ thereby plays a vital role in the parameter evolution, determining the amount by which the parameter values can vary from time $j$ to $j+1$.
For the time being, we assume that this matrix is time-invariant, such that $\mathsf{E}_{j+1} = \mathsf{E}$ for some prescribed matrix $\mathsf{E}$.

Using the parameter evolution model in \eqref{Eq:Param_model} allows us to treat the joint estimation problem in a similar manner to state estimation, by forming an augmented state vector
\[
z_j = \left[  \begin{array}{c} x_j \\ \theta_j \end{array} \right] \in \R^{d+p}
\]
with corresponding augmented state evolution equation
\[
Z_{j+1} = \widetilde{F}(Z_j) + \widetilde{V}_{j+1}, \quad \widetilde{V}_{j+1} \sim \mathcal{N}(0, \widetilde{\mathsf{C}})
\]
where 
\[
\tilde{F}(Z_j) = \left[  \begin{array}{c} F(X_j,\theta_j) \\ \theta_j \end{array} \right]  \quad \text{and} \quad \widetilde{\mathsf{C}} = \left[  \begin{array}{cc} \mathsf{C} & \mathsf{0} \\ \mathsf{0} & \mathsf{E} \end{array} \right] .
\]
The following algorithm outlines a particle filtering procedure to perform joint state and time-varying parameter estimation; we will refer to this algorithm as the PF-TVP algorithm (Particle Filter for Time-Varying Parameter Estimation).

\vspace{0.5cm}

\noindent\makebox[\linewidth]{\rule{\textwidth}{1pt}}\\[0.2cm]
\textbf{Algorithm:} Particle Filter for Time-Varying Parameter Estimation (PF-TVP) \\[0.2cm]
Given the joint prior density $\pi(x_0,\theta_0|D_0)$, draw an initial sample of state and time-varying parameter particles with equal weights:
\[
\mathcal{S}_0 = \big\{ \big(x_0^{(n)}, \theta_0^{(n)}, w_0^{(n)}\big)\big\}_{n=1}^N, \qquad w_0^{(1)} = \cdots = w_0^{(N)} = \frac{1}{N}
\]
Set $j=0$. While $j<T$, where $T$ is the number of available observations, do:
\begin{enumerate}
\item Propagate forward to compute the state predictors: 
\[
\widehat{x}_{j+1}^{(n)} = F\big(x_j^{(n)},\theta_j^{(n)}\big), \qquad n = 1, \dots, N
\]
\item Calculate the fitness weights for each $n$:
\[
g_{j+1}^{(n)} = w_j^{(n)}  \pi(y_{j+1} | \widehat{x}_{j+1}^{(n)}) , \qquad g_{j+1}^{(n)} \ \leftarrow \ \frac{g_{j+1}^{(n)} }{\sum_n g_{j+1}^{(n)}} 
\]
\item Draw with replacement the auxiliary indices $\ell_n \in \{1, 2, \dots, N\}$, $n=1,\dots, N$, using the fitness probabilities:
\[
P\{\ell_n = k \} = g_{j+1}^{(k)}
\]
\item Reshuffle current samples of states, time-varying parameters, and predictors:
\[
x_j^{(n)} \ \leftarrow \ x_j^{(\ell_n)} , \qquad \theta_j^{(n)} \ \leftarrow \ \theta_j^{(\ell_n)} , \qquad \widehat{x}_j^{(n)} \ \leftarrow \ \widehat{x}_j^{(\ell_n)} , \qquad n = 1, \dots, N
\]
\item Repropagate and innovate the reshuffled state particles:
\[
x_{j+1}^{(n)} = F\big(x_j^{(n)}, \theta_j^{(n)} \big) + v_{j+1}^{(n)}, \qquad v_{j+1}^{(n)}\sim\mathcal{N}(0,\mathsf{C}), \qquad n = 1, \dots, N
\]
\item Propagate and innovate the reshuffled time-varying parameter particles:
\[
\theta_{j+1}^{(n)} = \theta_j^{(n)} + \xi_{j+1}^{(n)}, \qquad \xi_{j+1}^{(n)}\sim\mathcal{N}(0,\mathsf{E}), \qquad n = 1, \dots, N
\]
\item Recalculate the weights for each $n$:
\[
w_{j+1}^{(n)} = \Frac{ \pi(y_{j+1} | x_{j+1}^{(n)} ) }{\pi(y_{j+1} | \widehat{x}_{j+1}^{(n)}) }, \qquad w_{j+1}^{(n)} \ \leftarrow \ \frac{w_{j+1}^{(n)} }{\sum_n w_{j+1}^{(n)}} 
\]
\item Set $j \leftarrow j+1$ and repeat from Step 1.
\end{enumerate}
\noindent\makebox[\linewidth]{\rule{\textwidth}{1pt}}\\

\vspace{0.5cm}

There are a number of significant implementation points to carefully consider before applying the PF-TVP algorithm.  
In addition to selecting a reasonable prior distribution on the initial values of the states and parameters (and assuming the availability of sequential observations of the system states), a practitioner must specify the following in advance of running the algorithm:
\begin{itemize}
\item The sample size $N$ of particles used in the Monte Carlo approximation;
\item The covariance matrices of the state evolution-observation model noise processes; i.e., the matrices $\mathsf{C}$ and $\mathsf{D}$ in \eqref{Eq:State_Evolution} and \eqref{Eq:Observation};
\item The covariance matrix of the parameter evolution model noise process; i.e., the matrix $\mathsf{E}$ in \eqref{Eq:Param_model};
\item The numerical time propagation scheme and corresponding time step.
\end{itemize}
The goal in choosing an appropriate sample size $N$ is to have a large enough number of particles so that the corresponding Monte Carlo approximation is able to well represent the underlying probability distributions; this can vary by problem and may require preliminary testing and/or validation for the problem at hand.
For some suggested approaches in the literature, see, e.g., \cite{Slivinski2016, Fox2003}.
As one might reasonably expect, increasing $N$ increases the computational cost of running the PF-TVP algorithm, with the bulk of this cost relating to the additional time integration steps in solving the ODE model (2 $\times$ the number of time steps for each additional particle).

The choice of the prescribed noise covariance matrices $\mathsf{C}$ and $\mathsf{D}$ play an important role in the filter's trust of the ODE model output and available data, respectively.
While each component could generally have a different variance, in this work we assume that these covariance matrices have the form
\[
\mathsf{C} = \sigma_C^2 \mathsf{I}_d \qquad \text{and} \qquad \mathsf{D} = \sigma_D^2 \mathsf{I}_m
\]
for some fixed constants $\sigma_C$ and $\sigma_D$, where $\mathsf{I}_d$ denotes the $d\times d$ identity matrix (and $d$ is the dimension of the state vector) and, similarly, $\mathsf{I}_m$ is the $m\times m$ identity (and $m$ is the dimension of the observation vector).
How to systematically assign these matrices is a topic of interest, with various approaches suggested in the literature; for example, the work in \cite{Ozkan2013} proposes an adaptive particle filtering scheme to sequentially estimate the main diagonal entries of these matrices (assuming diagonal forms).
In the setting of ODEs, the work in \cite{Arnold2013} describes an approach to systematically assign $\mathsf{C}$ at each time for each particle based on classic numerical error estimates for ODE solvers.

When implementing the PF-TVP algorithm, assigning the covariance matrix $\mathsf{E}$ in the parameter evolution model \eqref{Eq:Param_model} plays a vital role in the parameter tracking process.
To begin, we assume the form
\[
 \mathsf{E} = \sigma_E^2 \mathsf{I}_p 
\]
with some fixed constant $\sigma_E$, where $\sigma_E^2$ is the parameter noise (or drift) variance, $\mathsf{I}_p$ denotes the $p\times p$ identity matrix, and $p$ is the dimension of the time-varying parameter vector.
We will demonstrate the importance of this choice in the computed example that follows and further discuss its systematic assignment in the next section.

Since this works focuses on estimating time-varying parameters in dynamical systems modeled using ODEs, one of the most important underlying aspects of the proposed algorithm is the numerical solution to the ODE system \eqref{Eq:ODE_model}; in particular, this appears in Steps 1 and 5 of the PF-TVP algorithm, in propagating each particle forward from time $j$ to $j+1$ in each sequential iteration. 
Choosing an appropriate numerical technique to solve the ODE model is therefore a high priority, and in doing so, we must consider issues such as stiffness and selecting a time step size to guarantee stability, accuracy, and convergence of the numerical method.
For the simulations in this work, we utilize backward differentiation formula (BDF) methods, a class of implicit linear multistep methods specifically amenable to solving stiff problems \cite{LeVeque2007, Iserles2009}.
Using linear multistep methods to compute the forward particle propagation was shown to uphold the Markov property for state evolution in \cite{Arnold2013}.
To apply an $r$-step BDF method, we must store the $r$ past state approximations $x_{j-r+1}^{(n)}, \dots, x_j^{(n)}$, for each particle $n$, for use in the updating formula.
This thereby requires a reshuffling of the state histories along with the current state values in Step 4 of the algorithm.
To apply an $r$-step Adams method, note that we would instead need to store and reshuffle the past history of function values $f(t_{j-r+1}, x_{j-r+1}^{(n)}, \theta_{j-r+1}^{(n)}), \dots, f(t_{j}, x_{j}^{(n)}, \theta_{j}^{(n)})$.

In addition to the above, it is also important to consider the potential effects of sample degeneracy and impoverishment; see, e.g., \cite{Li2014}.
To this end, in the simulations that follow, we keep track of the number of unique particles retained at each resampling step of the filter, calculating the \textit{particle retention} as the ratio of unique auxiliary indices drawn in Step 3 over the total number of particles $N$.
Due to the independent nature of the forward propagation for each particle in Steps 1 and 5, we also note the availability of different implementation strategies for speeding up the computing time over a naive sequential implementation, including parallelizing the propagation steps.
In this work, we utilize a vectorized version such that all of the particles are propagated forward at once, in a similar style to \cite{Arnold2015}.
The following computed example demonstrates the performance of the PF-TVP algorithm using a fixed variance for the parameter drift.

\subsection{Computed Example: Forced Logistic Equation}
\label{Sec:Ex_ForcedLogistic}

As a first example, consider a forced logistic equation of the form
\begin{equation}\label{Eq:Forced_Logistic}
\frac{dx}{dt} = ax -bx^2 + \theta(t)
\end{equation}
where the state $x=x(t)$ represents the size of a population at time $t$, the growth rate $a$ and overcrowding parameter $b$ are fixed constants, and $\theta(t)$ is a time-varying harvesting or repopulation parameter; a similar system is described in \cite{Scarpello2008}. 

\begin{figure}[t!]
  \centerline{\includegraphics[width=0.95\textwidth]{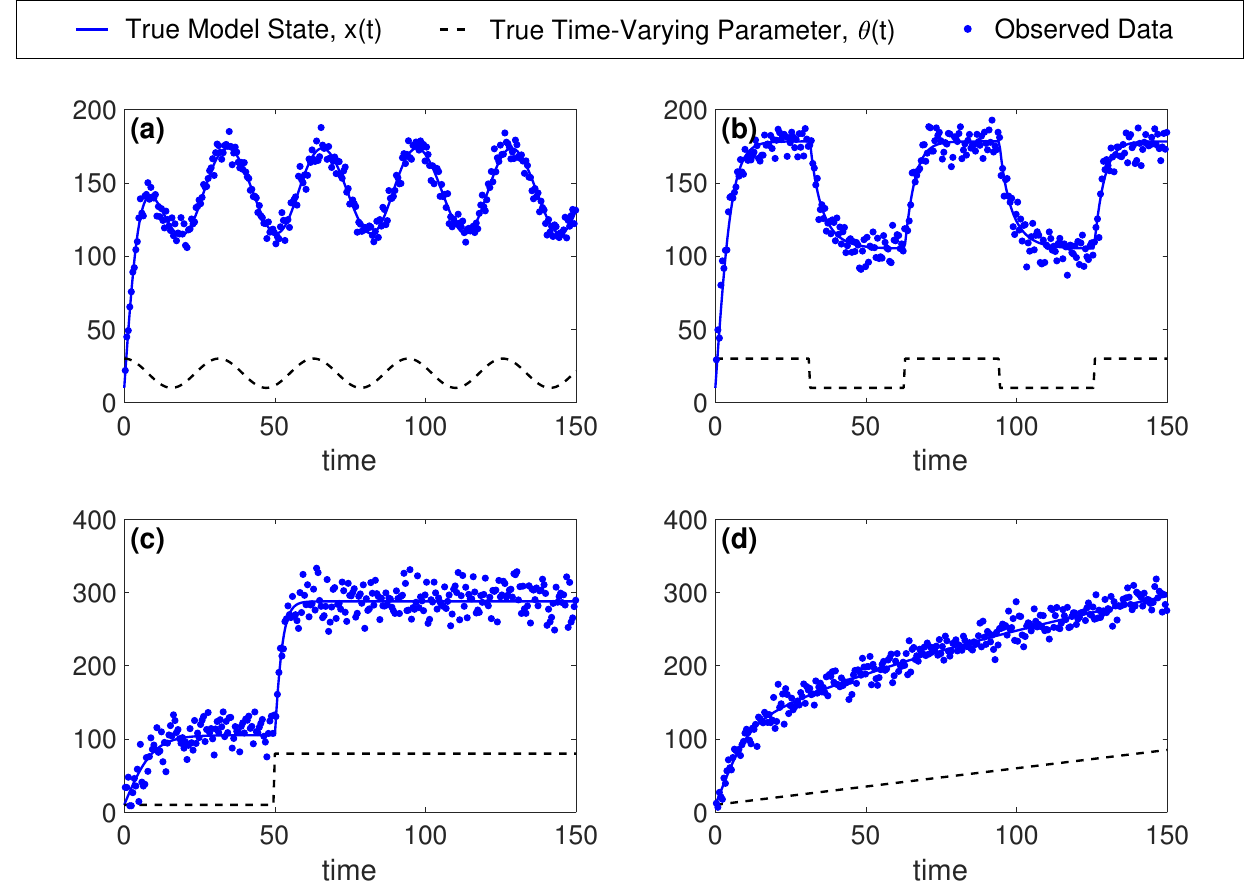} }
  \caption{Synthetic data generated from the forced logistic model in \eqref{Eq:Forced_Logistic} using four different time-varying parameters: (a) the sinusoidal function in \eqref{Eq:Logistic_Sinusoid}; (b) the multiple step function in \eqref{Eq:Logistic_ManySteps}; (c) the single step function in \eqref{Eq:Logistic_BigStep}; and (d) the linear function in \eqref{Eq:Logistic_Line}. In each plot, the true model state $x(t)$ is shown in solid blue, the true time-varying parameter $\theta(t)$ is shown in dashed black, and the observed data are shown in blue markers. }
  \label{Fig:Data_Logistic}
\end{figure}

To generate synthetic data from this equation, we assign the initial value $x(0) = 10$, constant parameter values $a = 0.01$, $b = 0.001$, and let $\theta(t)$ take the form of four different time-varying functions:
\begin{itemize}
\item[(a)] Sinusoidal function: 
\begin{equation}\label{Eq:Logistic_Sinusoid}
\theta(t) = 20 + 10\cos(0.2t)
\end{equation}
\item[(b)] Multiple step function:  
\begin{equation}\label{Eq:Logistic_ManySteps}
\theta(t) = 10s(t/10) + 20
\end{equation}
where $s(t)$ is a square wave function with period $2\pi$ that alternates between $-1$ and $1$
\item[(c)] Single step function:  
\begin{equation}\label{Eq:Logistic_BigStep}
\theta(t) = \begin{cases} 10, & 0\leq t<50 \\ 80, & 50\leq t \leq 150 \end{cases} 
\end{equation}
\item[(d)] Linear function:  
\begin{equation}\label{Eq:Logistic_Line}
\theta(t) = 0.5t + 10.
\end{equation}
\end{itemize}
We compute the true system states using MATLAB's \texttt{ode15s}, a variable-step solver that uses numerical differentiation formulas of orders 1-5 \cite{MATLAB_ODE}, taking observations every 0.5 time units over the interval $[0,150]$ and adding zero-mean Gaussian noise with standard deviation set to be 20\% of the standard deviation of the underlying state.
Figure~\ref{Fig:Data_Logistic} plots the true states, time-varying parameters, and noisy state observations for each case.

In our first simulation, we apply the PF-TVP algorithm to estimate $\theta(t)$ using the data shown in Figure~\ref{Fig:Data_Logistic}(a).
We use a sample of $N=1,000$ particles, set the state evolution and observation noise variances to $\mathsf{C} = \sigma_C^2$ and $\mathsf{D} = \sigma_D^2$, respectively, with $\sigma_C = 0.5$ and $\sigma_D = 10$, 
and use the two-step backward differentiation formula of order 2 (BDF2) for time propagation with step size 0.25.
Note that the step size was chosen such that the solver remains absolutely stable,
and results are consistent when using larger samples.
We draw the initial sample of state particles from a uniform prior distribution 0.5 to 1.5 times the initial value $x(0)$; similarly, we draw the initial sample of time-varying parameters from a uniform prior 0.5 to 1.5 times the value of $\theta(0)$.

\begin{figure}[t!]
  \centerline{\includegraphics[width=0.95\textwidth]{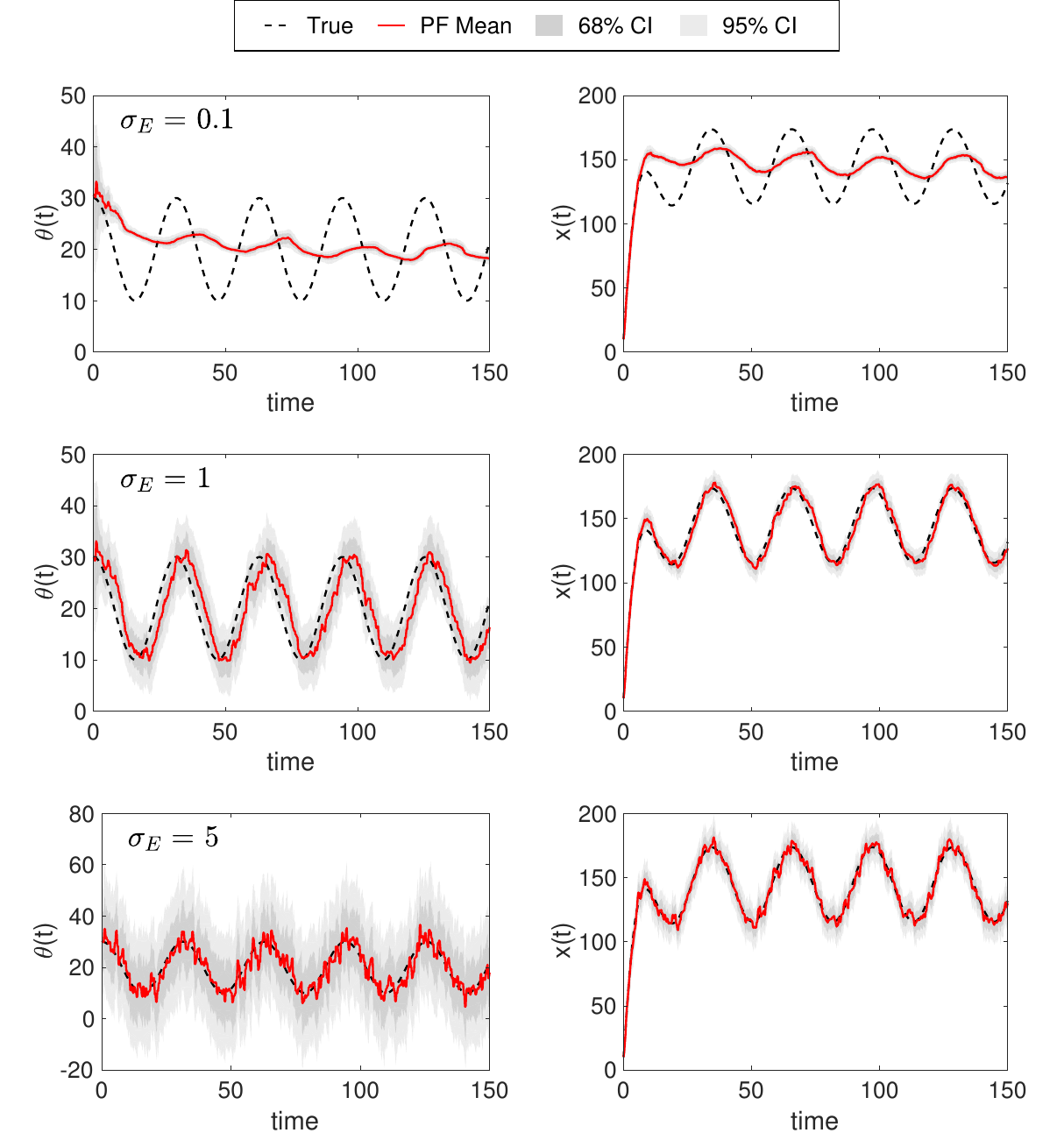} }
  \caption{Time-varying parameter estimates (left) and corresponding state estimates (right) obtained when applying the PF-TVP algorithm with the following fixed parameter drift standard deviations: (top row) $\sigma_E=0.1$; (middle row) $\sigma_E=1$; (bottom row) $\sigma_E=5$.  In each plot, the underlying true time-varying parameter $\theta(t)$ or state $x(t)$ is shown in dashed black, the PF-TVP mean is shown in solid red, and the 68\% and 95\% credible intervals are shown in dark and light gray, respectively.}
  \label{Fig:Results_Logistic_FixedSigma}
\end{figure}

Figure~\ref{Fig:Results_Logistic_FixedSigma} shows the resulting $\theta(t)$ estimates and corresponding $x(t)$ estimates obtained when using three different fixed values for the standard deviation of the parameter evolution noise process in \eqref{Eq:Param_model}: $\sigma_E = 0.1$, $\sigma_E = 1$, and $\sigma_E = 5$, where $\mathsf{E} = \sigma_E^2$.
Note that when $\sigma_E = 0.1$, the small variance restricts the movement of the parameter sample, causing the resulting $\theta(t)$ estimate to be notably damped down compared the true behavior of the time-varying parameter; this effect is also seen in the corresponding estimate of $x(t)$.  
In this case, the credible intervals for both the parameter and state estimates are very close around the sample means and do not well represent the underlying truth.
When $\sigma_E = 1$, the increased variance allows the parameter sample to track fairly closely the underlying true $\theta(t)$, with a small time delay in capturing the oscillations; however, the true dynamics are mostly captured within the credible intervals.
Results hold similarly for the corresponding estimate of $x(t)$.
While the algorithm is still able to track the underlying $\theta(t)$ and corresponding $x(t)$ when $\sigma_E = 5$, the sample means and credible intervals are much wider around the truth, indicating considerable uncertainty in the mean estimates.  
These results demonstrate the clear impact that the choice of $\sigma_E$ has on the time-varying parameter estimate, as well as the corresponding estimate of the model state.
This leads us in the next section to consider a modification of the PF-TVP algorithm that allows for online estimation of $\sigma_E$ during the filtering process.



\section{Catching the Drift: Enhanced Parameter Tracking with Online Variance Estimation}
\label{Sec:PF_TVP_Enhanced}

As demonstrated in the previous example, the PF-TVP algorithm relies on a fixed value for the parameter drift standard deviation $\sigma_E$ being set as an input before running the filtering procedure.  
The results in Figure~\ref{Fig:Results_Logistic_FixedSigma} show how significantly the choice of this parameter can affect the resulting time-varying parameter estimate.
However, in most cases, it is not clear how to choose an appropriate value for $\sigma_E$ in advance of running the algorithm, and tuning over several runs of the filter may be necessary in order to obtain reasonable results.
To overcome this limitation, we propose here a modification to enhance the PF-TVP algorithm by providing a systematic approach to simultaneously estimate $\sigma_E$ along with the time-varying parameter of interest.
Assuming the drift variance is time-invariant, we invoke the kernel smoothing approach described in \cite{LiuWest2001} for combined state and static parameter estimation, where here $\sigma_E$ is the static parameter to estimate; we will refer to $\sigma_E$ as the \textit{drift constant} moving forward.

Let $z_j = (x_j, \theta_j)$ denote the augmented vector jointly representing the model states and time-varying parameters, and for ease of notation, let $\sigma = \sigma_E$ represent the drift constant.
From Bayes' theorem, it follows that
\begin{eqnarray*}
\pi(z_{j+1},\sigma | D_{j+1}) &\propto& \pi(y_{j+1} | z_{j+1}, \sigma) \pi(z_{j+1}, \sigma | D_j) \\[0.2cm]
\ &\propto& \pi(y_{j+1} | z_{j+1}, \sigma) \pi(z_{j+1} | \sigma, D_j) \pi(\sigma | D_j) 
\end{eqnarray*}
which directly depends on the drift constant distribution $\pi(\sigma | D_j)$ in the updating formula.
To approximate $\pi(\sigma | D_j)$, we assume a Monte Carlo sample $\{\sigma_j^{(n)},w_j^{(n)}\}_{n=1}^N$, with mean and covariance at time $j$ given by
\[
\bar{\sigma}_{j} = \sum_{n=1}^N w_{j}^{(n)} \sigma_{j}^{(n)} \qquad \text{and} \qquad \mathsf{S}_{j} = \sum_{n=1}^N w_{j}^{(n)} \big(\sigma_{j}^{(n)}-\bar{\sigma}_{j}\big) \big(\sigma_{j}^{(n)}-\bar{\sigma}_{j}\big)^\mathsf{T}
\]
respectively, and use a Gaussian mixture model of the type suggested in \cite{LiuWest2001}. 
Cast in the setting of artificially evolving $\sigma$, this gives
\[
\pi(\sigma_{j+1} | D_j) \approx \sum_{n=1}^n w_j^{(n)} \mathcal{N}( \sigma_{j+1} | \widehat{\sigma}_j^{(n)}, h^2\mathsf{S}_j )
\]
where
\[
\widehat{\sigma}_j^{(n)} = a\sigma_j^{(n)} + (1-a)\bar{\sigma}_j, \qquad a = \frac{3\delta-1}{2\delta}
\]
shrinks each $\sigma_j^{(n)}$ value towards the sample mean $\bar{\sigma}_{j}$ by the factor $0 < a < 1$ as defined, with specified discount factor $0 < \delta < 1$, and $h^2 = 1 - a^2$. 
These factors are chosen in such a way to avoid the loss of information commonly observed in artificial evolution of static parameters; for more details, we refer to \cite{LiuWest2001, West1993a, West1993b}.

This idea can be embedded into the PF-TVP algorithm to perform sequential estimation of the drift constant $\sigma$ and thereby improve the time-varying parameter estimate.
The following algorithm outlines the modified particle filtering procedure to perform joint state and time-varying parameter estimation, along with online estimation of the parameter drift constant; we will refer to this as the PF-TVP+ algorithm (Enhanced Particle Filter for Time-Varying Parameter Estimation).

\vspace{0.5cm}

\noindent\makebox[\linewidth]{\rule{\textwidth}{1pt}}\\[0.2cm]
\textbf{Algorithm:} Enhanced Particle Filter for Time-Varying Parameter Estimation (PF-TVP+) \\[0.2cm]
(with Online Estimation of the Parameter Drift Constant)\\[0.2cm]
Given the joint prior density $\pi(x_0,\theta_0,\sigma_0|D_0)$, draw an initial sample of state, time-varying parameter, and drift constant particles with equal weights:
\[
\mathcal{S}_0 = \big\{ \big(x_0^{(n)}, \theta_0^{(n)}, \sigma_0^{(n)}, w_0^{(n)}\big)\big\}_{n=1}^N, \qquad w_0^{(1)} = \cdots = w_0^{(N)} = \frac{1}{N}
\]
Calculate the sample mean and covariance of the drift constants:
\[
\bar{\sigma}_0 = \sum_{n=1}^N w_0^{(n)} \sigma_0^{(n)}, \qquad \mathsf{S}_0 = \sum_{n=1}^N w_0^{(n)} \big(\sigma_0^{(n)}-\bar{\sigma}_0\big) \big(\sigma_0^{(n)}-\bar{\sigma}_0\big)^\mathsf{T}
\]
Set $j=0$. While $j<T$, where $T$ is the number of available observations, do:
\begin{enumerate}
\item Shrink the drift constant particles toward the sample mean for each $n$:
\[
\widehat{\sigma}_j^{(n)} = a\sigma_j^{(n)} + (1-a)\bar{\sigma}_j, \qquad a = \frac{3\delta-1}{2\delta}
\]
with specified discount factor $0 < \delta < 1$
\item Propagate forward to compute the state predictors: 
\[
\widehat{x}_{j+1}^{(n)} = F\big(x_j^{(n)},\theta_j^{(n)}\big), \qquad n = 1, \dots, N
\]
\item Calculate the fitness weights for each $n$:
\[
g_{j+1}^{(n)} = w_j^{(n)}  \pi(y_{j+1} | \widehat{x}_{j+1}^{(n)}) , \qquad g_{j+1}^{(n)} \ \leftarrow \ \frac{g_{j+1}^{(n)} }{\sum_n g_{j+1}^{(n)}} 
\]
\item Draw with replacement the auxiliary indices $\ell_n \in \{1, 2, \dots, N\}$, $n=1,\dots, N$, using the fitness probabilities:
\[
P\{\ell_n = k \} = g_{j+1}^{(k)}
\]
\item Reshuffle current samples of states, time-varying parameters, drift constants, and predictors:
\[
x_j^{(n)} \ \leftarrow \ x_j^{(\ell_n)} , \qquad \theta_j^{(n)} \ \leftarrow \ \theta_j^{(\ell_n)} , \qquad \widehat{\sigma}_j^{(n)} \ \leftarrow \ \widehat{\sigma}_j^{(\ell_n)} , \qquad \widehat{x}_j^{(n)} \ \leftarrow \ \widehat{x}_j^{(\ell_n)} , \qquad n = 1, \dots, N
\]
\item Repropagate and innovate the reshuffled state particles:
\[
x_{j+1}^{(n)} = F\big(x_j^{(n)}, \theta_j^{(n)} \big) + v_{j+1}^{(n)}, \qquad v_{j+1}^{(n)}\sim\mathcal{N}(0,\mathsf{C}), \qquad n = 1, \dots, N
\]
\item Artificially evolve the reshuffled drift constant particles:
\[
\sigma_{j+1}^{(n)} = \widehat{\sigma}_j^{(n)} + \zeta_{j+1}^{(n)}, \qquad  \zeta_{j+1}^{(n)}\sim\mathcal{N}(0,h^2\mathsf{S}_j), \qquad \qquad n = 1, \dots, N
\]
where $h^2 = 1-a^2$
\item Form the drift covariance matrices using the updated drift constants:
\[
\mathsf{E}_{j+1}^{(n)} = \text{diag}\big((\sigma_{j+1}^{(n)})^2\big), \qquad n = 1, \dots, N
\]
\item Propagate and innovate the reshuffled time-varying parameter particles:
\[
\theta_{j+1}^{(n)} = \theta_j^{(n)} + \xi_{j+1}^{(n)}, \qquad \xi_{j+1}^{(n)}\sim\mathcal{N}(0,\mathsf{E}_{j+1}^{(n)}), \qquad n = 1, \dots, N
\]
\item Recalculate the weights for each $n$:
\[
w_{j+1}^{(n)} = \Frac{ \pi(y_{j+1} | x_{j+1}^{(n)} ) }{\pi(y_{j+1} | \widehat{x}_{j+1}^{(n)}) }, \qquad w_{j+1}^{(n)} \ \leftarrow \ \frac{w_{j+1}^{(n)} }{\sum_n w_{j+1}^{(n)}} 
\]
\item Recalculate the sample mean and covariance of the drift constants:
\[
\bar{\sigma}_{j+1} = \sum_{n=1}^N w_{j+1}^{(n)} \sigma_{j+1}^{(n)}, \qquad \mathsf{S}_{j+1} = \sum_{n=1}^N w_{j+1}^{(n)} \big(\sigma_{j+1}^{(n)}-\bar{\sigma}_{j+1}\big) \big(\sigma_{j+1}^{(n)}-\bar{\sigma}_{j+1}\big)^\mathsf{T}
\]
\item Set $j \leftarrow j+1$ and repeat from Step 1.
\end{enumerate}
\noindent\makebox[\linewidth]{\rule{\textwidth}{1pt}}\\

\vspace{0.5cm}

The enhanced PF-TVP+ algorithm differs from PF-TVP mainly in Steps 1, 7, and 8; i.e., in shrinking, propagating, and using the updated $\sigma_{j+1}^{(n)}$ values to form the parameter drift covariance matrix $\mathsf{E}_{j+1}^{(n)}$.
Note that $\mathsf{E}_{j+1}^{(n)}$ is now individually assigned for each particle $n$ at each time, based on the current estimate of $\sigma_{j+1}^{(n)}$, as opposed to being set a priori to a constant matrix $\mathsf{E}$ and utilized for all particles at all times throughout the filtering procedure.
The notation used in Step 8 is meant to reflect two possible scenarios: (i) when a single constant drift constant is estimated for each of the unknown time-varying parameters in a system, so that $\sigma_{j+1}^{(n)}\in\R$ and
\[
\mathsf{E}_{j+1}^{(n)} \ = \ \big(\sigma_{j+1}^{(n)}\big)^2 \ \mathsf{I}_p \ = \ \left[ \begin{array}{ccc} \big(\sigma_{j+1}^{(n)}\big)^2 & \ & \ \\ \ & \ddots & \ \\ \ & \ & \big(\sigma_{j+1}^{(n)}\big)^2  \end{array} \right]_{p\times p}
\]
where $p$ is the dimension of the time-varying parameter vector; and (ii) when individual drift constants are estimated for each time-varying parameter, so that $\sigma_{j+1}^{(n)}\in\R^p$ and
\[
\mathsf{E}_{j+1}^{(n)} \ = \ \left[ \begin{array}{ccc} \big(\sigma_{j+1,1}^{(n)}\big)^2 & \ & \ \\ \ & \ddots & \ \\ \ & \ & \big(\sigma_{j+1,p}^{(n)}\big)^2  \end{array} \right]_{p\times p}
\]
where $\sigma_{j+1,i}^{(n)}$ denotes the $i$th entry of the vector, $i=1,\dots,p$.
In the simulations that follow, we also apply an inverse logit transformation in generating the $\sigma_{j+1}^{(n)}$ values at each step, in order to keep the samples between a small choice of $\sigma_\text{min}>0$ and reasonably large choice of $\sigma_\text{max}$.
A similar approach was taken in \cite{Arnold2013} when estimating static parameters with prescribed lower and upper bounds.

In the computed examples that follow, we demonstrate the capability of the PF-TVP+ algorithm in estimating the parameter drift constant(s) along with the time-varying parameter(s) of interest and the corresponding system state(s) in two ODE models: the forced logistic equation, and a forced harmonic oscillator with two time-varying parameters.

\subsection{Computed Example: Forced Logistic Equation}
\label{Sec:Ex_Logistic_Enhanced}

We return to the forced logistic model described in Section~\ref{Sec:Ex_ForcedLogistic} and consider the same data shown in Figure~\ref{Fig:Data_Logistic}.
Our aim is now to apply the PF-TVP+ algorithm to estimate $\sigma_E$ along with $\theta(t)$ and $x(t)$ for each of the four data sets.
In each case, we initialize the filter in the same manner as before, with a sample size of $N=1,000$ particles, $\sigma_C = 0.5$, $\sigma_D = 10$, 
and use BDF2 for time propagation with step size 0.25.
In addition, we set $\sigma_\text{min}=0.05$, $\sigma_\text{max} = 10$, and draw an initial sample of drift constants using an inverse logit transformation so that $\{\sigma_{0}^{(n)}\}_{n=1}^N$ is approximately uniformly distributed between $\sigma_\text{min}$ and $\sigma_\text{max}$.
We use a discount factor of $\delta = 0.96$, which corresponds to the shrinkage parameters $a\approx 0.98$ and $h \approx 0.2$.

\begin{figure}[t!]
  \centerline{\includegraphics[width=0.95\textwidth]{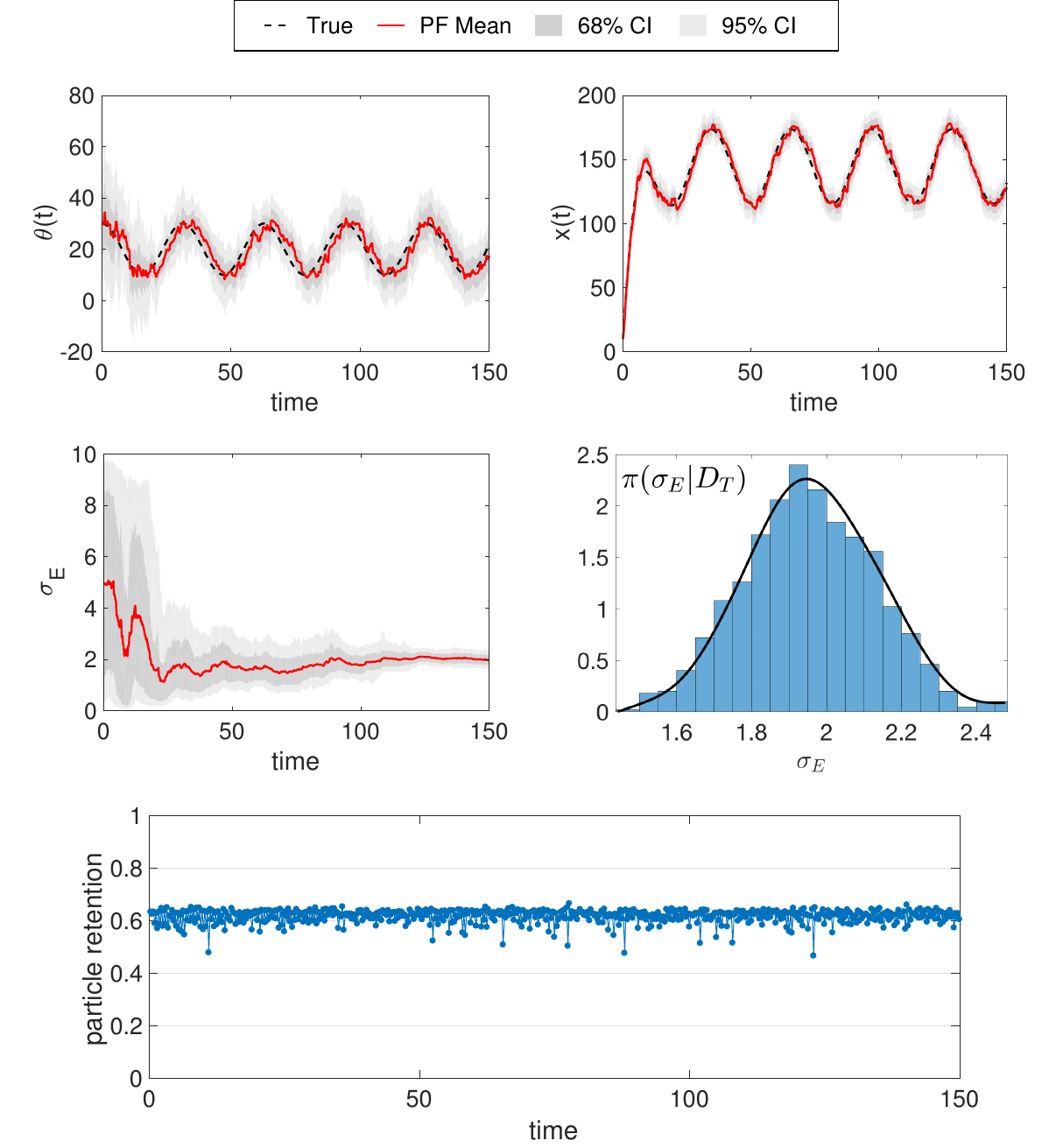}  }
  \caption{Results using the PF-TVP+ algorithm to estimate the sinusoidal $\theta(t)$ in the forced logistic model \eqref{Eq:Forced_Logistic} given the data in Figure~\ref{Fig:Data_Logistic}(a). Top row: Time-varying parameter estimate (left) and corresponding state estimate (right). In each plot, the underlying true time-varying parameter $\theta(t)$ or state $x(t)$ is shown in dashed black, the PF-TVP+ mean is shown in solid red, and the 68\% and 95\% credible intervals are shown in dark and light gray, respectively. Middle row: Time series estimate for the drift constant $\sigma_E$ (left) and histogram of the posterior sample at time $t=150$ (right). Bottom row: Particle retention at each time step.}
  \label{Fig:Results_Logistic_Enhanced_Sinusoid}
\end{figure}

Figure~\ref{Fig:Results_Logistic_Enhanced_Sinusoid} shows the PF-TVP+ results for the data generated with the sinusoidal $\theta(t)$ shown in Figure~\ref{Fig:Data_Logistic}(a).
Note that the time series estimate for $\sigma_E$ converges to a mean value of approximately 1.95 at time $t=150$, with 95\% of the posterior sample between values of approximately 1.62 and 2.29, as seen in the corresponding histogram.
The estimates of $\theta(t)$ and $x(t)$ are able to well track the underlying truth, with the particle retention staying around 60\% at most time steps, dropping down to 46.6\% at the lowest.

\begin{figure}[t!]
  \centerline{\includegraphics[width=1.1\textwidth]{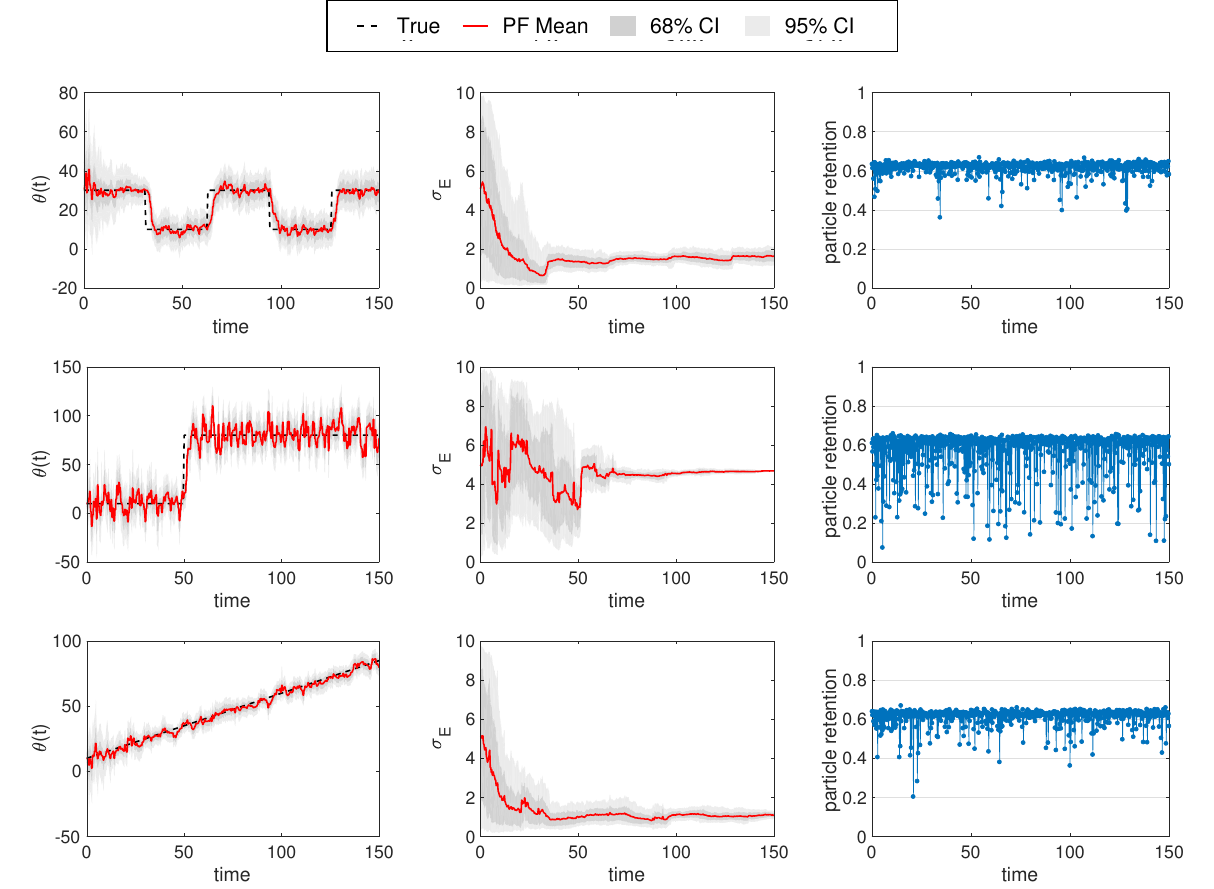} }
  \caption{Results using the PF-TVP+ algorithm to estimate $\theta(t)$ in the forced logistic model \eqref{Eq:Forced_Logistic} given the following data sets: (top row) data generated with the multiple step $\theta(t)$ shown in Figure~\ref{Fig:Data_Logistic}(b); (middle row) data generated with the single step $\theta(t)$ shown in Figure~\ref{Fig:Data_Logistic}(c); (bottom row) data generated with the linear $\theta(t)$ shown in Figure~\ref{Fig:Data_Logistic}(d). Each row displays the corresponding time series estimates for $\theta(t)$ (left) and $\sigma_E$ (middle), along with the particle retention (right).}
  \label{Fig:Results_Logistic_Enhanced_Other}
\end{figure}

Figure~\ref{Fig:Results_Logistic_Enhanced_Other} gives the resulting $\theta(t)$, $x(t)$, and $\sigma_E$ estimates obtained by applying the PF-TVP+ algorithm to the remaining three data sets shown in Figure~\ref{Fig:Data_Logistic}(b)--(d).
Despite a small time lag in capturing the step increases and decreases, the algorithm is able to well track $\theta(t)$ overall when the true time-varying parameter is the multiple step function defined in \eqref{Eq:Logistic_ManySteps}. 
The estimate of $\sigma_E$ converges to a mean value around 1.63, with 95\% of the posterior sample between around 1.17 and 2.10.
While not shown, the corresponding estimate of $x(t)$ is also well captured.
However, the algorithm has more difficulty estimating $\sigma_E$ for the single step $\theta(t)$ in \eqref{Eq:Logistic_BigStep}, which involves a more drastic change in magnitude for the parameter value.
The $\sigma_E$ converges very tightly to a mean estimate of 4.67, and while the filter is still able to well capture the jump in magnitude of $\theta(t)$ at time $t=50$, the mean parameter estimate is less certain around the underlying truth. 
This could be due to the sample retaining very few unique particles, with the retention dropping at its lowest point to 7.4\%.
The algorithm is better able to track the linear $\theta(t)$ in \eqref{Eq:Logistic_Line}, which involves a similar but more steady increase in magnitude for the parameter.
The particles are better retained in this case, and both the $\sigma_E$ and $\theta(t)$ estimates are refined over time, with the uncertainty around the mean visibly decreasing as the data are assimilated.

\subsection{Computed Example: Forced Harmonic Oscillator}
\label{Sec:Ex_Oscillator_Enhanced}

As a second example, consider a forced harmonic oscillator of the form
\[
m p'' + b p' + k(t) p = q(t)
\]
which is often used in modeling the dynamics of a mass on a spring, where $p = p(t)$ represents the position of the mass at time $t$; see, e.g., \cite{Boyce_book}.  Letting $v = p'$ denote the velocity of the mass and setting $m=1$, we arrive at the following system of first-order differential equations:
\begin{eqnarray}
\frac{dp}{dt} &=& v \label{Eq:Forced_OscillatorP} \\[2mm]
\frac{dv}{dt} &=& -k(t) p - b v + q(t)  \label{Eq:Forced_OscillatorV}
\end{eqnarray}
where $b$ is a constant damping coefficient, $k(t)$ is a time-varying parameter relating to spring stiffness, and $q(t)$ represents time-varying external forcing.  
While spring stiffness is often assumed to be constant, time-varying stiffness can arise, e.g., in models of gear systems \cite{Sika2008, Shen2006}.
We consider several different cases below, in which we estimate the time-varying external forcing and stiffness parameters, $q(t)$ and $k(t)$, respectively.
Note that $q(t)$ has an additive relationship with the state variables in \eqref{Eq:Forced_OscillatorV}, while $k(t)$ is multiplicative.

\begin{figure}[t!]
  \centerline{\includegraphics[width=0.95\textwidth]{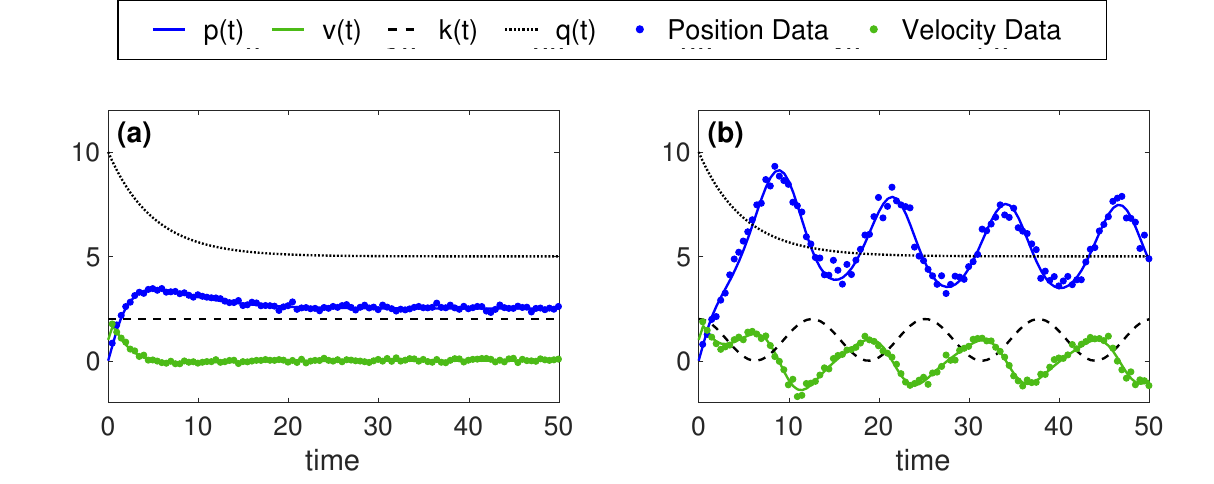} }
  \caption{Synthetic data generated from the forced harmonic oscillator in system \eqref{Eq:Forced_OscillatorP}--\eqref{Eq:Forced_OscillatorV} using damping coefficient $b=5$, time-varying external forcing $q(t) = 5\exp(-0.2t)+5$, and two different cases for the stiffness parameter $k(t)$: (a) constant stiffness with $k(t) = 2$; and (b) time-varying stiffness with $k(t) = 1+\cos(0.5t)$. In each plot, the true model state for position $p(t)$ is shown in solid blue and velocity $v(t)$ is shown in solid green, the true time-varying external forcing parameter $q(t)$ is shown in dotted black and stiffness $k(t)$ is shown in dashed black, and the observed position and velocity data are shown in blue and green markers, respectively. }
  \label{Fig:Data_Oscillator}
\end{figure}

To generate synthetic data from this system, we set the initial state values as $p(0) = 0$ and $v(0) = 1$, fix the damping coefficient to $b = 5$, and model the time-varying external forcing as an exponentially decaying function with $q(t) = 5\exp(-0.2t)+5$.
We consider two different functions for the stiffness parameter: (a) constant stiffness, with $k(t) = 2$; and (b) time-varying stiffness, with $k(t) = 1+\cos(0.5t)$.
As before, we compute the true system states using MATLAB's \texttt{ode15s} and take observations every 0.5 time units over the interval $[0,50]$.
Observations are corrupted by Gaussian noise with zero mean and standard deviation set to be 20\% of the standard deviation of the underlying system states.
Figure~\ref{Fig:Data_Oscillator} plots the true states, time-varying parameters, and noisy state observations for each case.
In the simulations that follow, we assume that the damping coefficient is known and fixed to $b=5$, and we set $\sigma_\text{min}=0.05$ and $\sigma_\text{max} = 5$.

\subsubsection*{Case 1: Estimating Time-Varying External Forcing}

In this case, given the data set shown in Figure~\ref{Fig:Data_Oscillator}(a) and assuming that the stiffness parameter $k(t)=2$ is known and fixed, our goal is to estimate the time-varying external forcing parameter $q(t)$.
We initialize the filter similarly as before, where here the state vector is given by $x(t) = [p(t); v(t)]\in\R^2$, so we set the state evolution and observation noise covariance matrices to $\mathsf{C} = \sigma_C^2 \mathsf{I}_2$ and $\mathsf{D} = \sigma_D^2 \mathsf{I}_2$, respectively, with $\sigma_C = 0.2$ and $\sigma_D = 0.2$.

\begin{figure}[t!]
  \centerline{\includegraphics[width=0.95\textwidth]{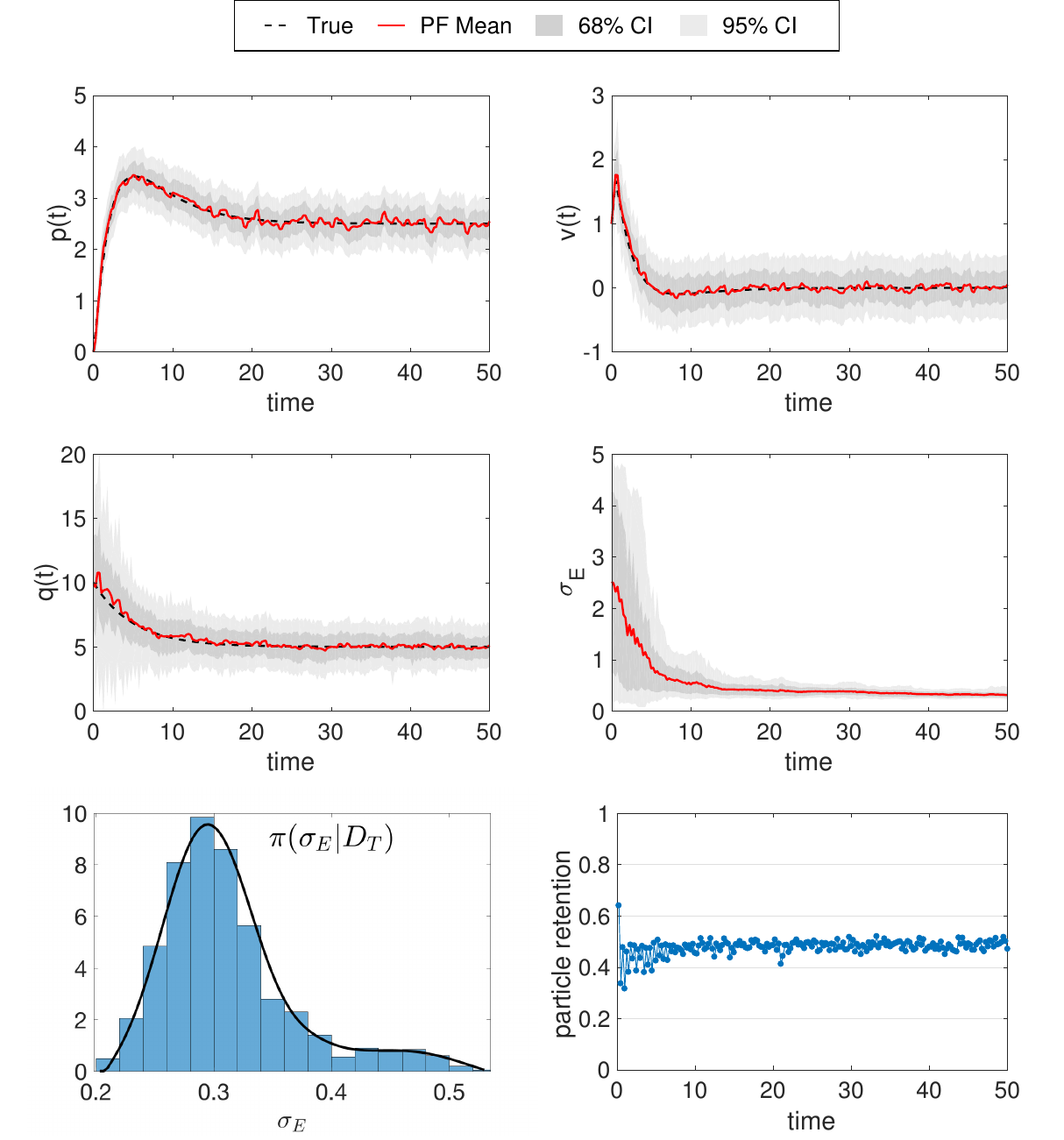} }
  \caption{Results using the PF-TVP+ algorithm to estimate the time-varying external forcing parameter $q(t)$ in system \eqref{Eq:Forced_OscillatorP}--\eqref{Eq:Forced_OscillatorV} given the data in Figure~\ref{Fig:Data_Oscillator}(a). Top row: State estimates for $p(t)$ (left) and $v(t)$ (right). Middle row: Time-varying parameter estimate for $q(t)$ (left) and time series estimate for the drift constant $\sigma_E$ (right). Bottom row: Histogram of the posterior sample for $\sigma_E$ at time $t=50$ (left) and particle retention at each time step (right).}
  \label{Fig:Results_Oscillator_Enhanced_Forcing}
\end{figure}

Figure~\ref{Fig:Results_Oscillator_Enhanced_Forcing} shows the resulting PF-TVP+ estimates for the two model states, forcing parameter $q(t)$, and drift constant $\sigma_E$, along with a plot of the particle retention.
Here we see that the filter is able to well track the underlying $q(t)$, with uncertainty around the mean decreasing over time as the estimate of $\sigma_E$ is refined.
The posterior mean estimate for $\sigma_E$ in this case is approximately 0.31, with about 95\% of the sample between values of 0.23 and 0.47.
After an initial drop, the particle retention stays around 50\%.

\subsubsection*{Case 2: Estimating Time-Varying Stiffness}

\begin{figure}[t!]
  \centerline{\includegraphics[width=0.95\textwidth]{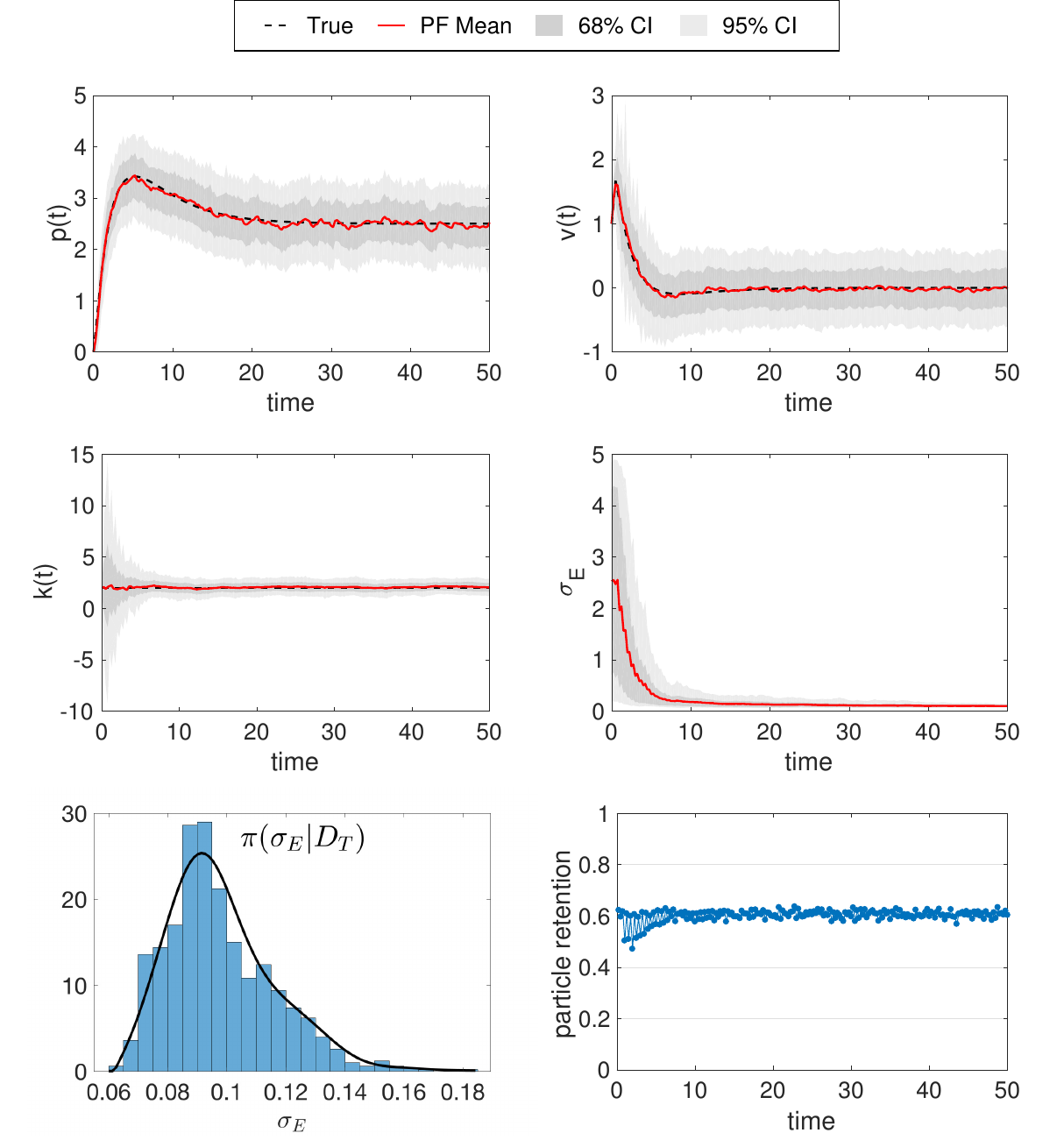}}
  \caption{Results using the PF-TVP+ algorithm to estimate the time-varying stiffness parameter $k(t)$ in system \eqref{Eq:Forced_OscillatorP}--\eqref{Eq:Forced_OscillatorV} given the data in Figure~\ref{Fig:Data_Oscillator}(a). Top row: State estimates for $p(t)$ (left) and $v(t)$ (right). Middle row: Time-varying parameter estimate for $k(t)$ (left) and time series estimate for the drift constant $\sigma_E$ (right). Bottom row: Histogram of the posterior sample for $\sigma_E$ at time $t=50$ (left) and particle retention at each time step (right).}
  \label{Fig:Results_Oscillator_Enhanced_ConstStiff}
\end{figure}

In the next case, we assume that the external forcing parameter $q(t)$ is known and instead aim to estimate the time-varying stiffness parameter $k(t)$.
Note that since $k(t)$ is a multiplicative parameter, its value affects the Jacobian matrix of the system \eqref{Eq:Forced_OscillatorP}--\eqref{Eq:Forced_OscillatorV}.
We start with the data set shown in Figure~\ref{Fig:Data_Oscillator}(a), where the stiffness is constant, and initialize the filter in a similar way as before, with $\sigma_C = 0.2$ and $\sigma_D = 0.5$.
Figure~\ref{Fig:Results_Oscillator_Enhanced_ConstStiff} shows the resulting PF-TVP+ estimates for the two model states, stiffness parameter $k(t)$, drift constant $\sigma_E$, and particle retention.
As seen in these plots, the algorithm is able to both recognize and well estimate the constant $k(t)$, with the uncertainty in the mean parameter estimate decreasing as the value of $\sigma_E$ converges to a posterior mean estimate of about 0.097. 
The relatively small $\sigma_E$ estimate reflects the algorithm's ability to detect a constant parameter and decrease the variance in the drift term accordingly.

\begin{figure}[t!]
  \centerline{\includegraphics[width=0.95\textwidth]{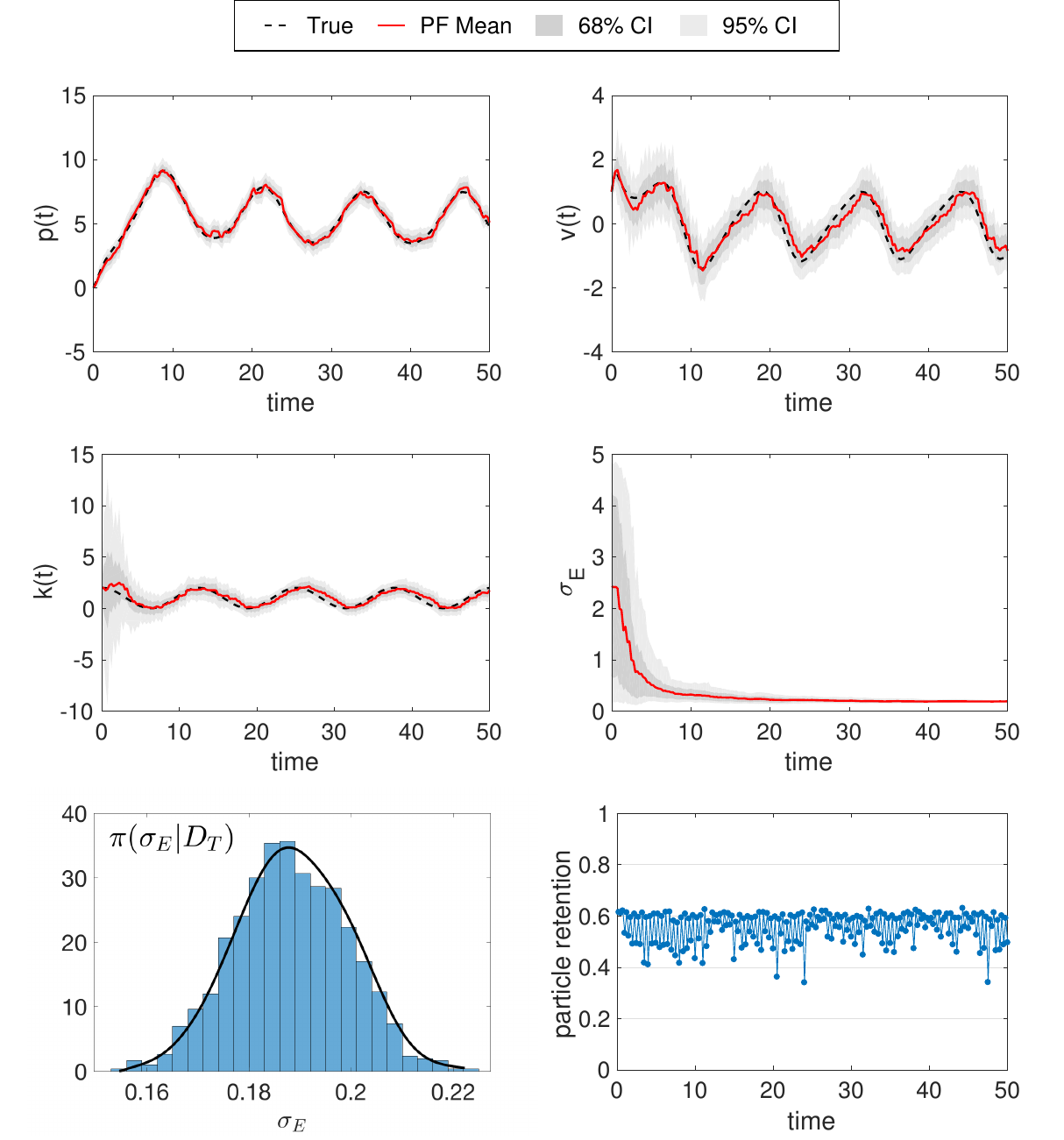} }
  \caption{Results using the PF-TVP+ algorithm to estimate the time-varying stiffness parameter $k(t)$ in system \eqref{Eq:Forced_OscillatorP}--\eqref{Eq:Forced_OscillatorV} given the data in Figure~\ref{Fig:Data_Oscillator}(b). Top row: State estimates for $p(t)$ (left) and $v(t)$ (right). Middle row: Time-varying parameter estimate for $k(t)$ (left) and time series estimate for the drift constant $\sigma_E$ (right). Bottom row: Histogram of the posterior sample for $\sigma_E$ at time $t=50$ (left) and particle retention at each time step (right).}
  \label{Fig:Results_Oscillator_Enhanced_TVStiff}
\end{figure}

Next, we use the data set shown in Figure~\ref{Fig:Data_Oscillator}(b), where the underlying stiffness parameter is sinusoidal.
Figure~\ref{Fig:Results_Oscillator_Enhanced_TVStiff} shows the resulting PF-TVP+ estimates, which demonstrate the filter's ability in capturing the behavior of $k(t)$ even in the challenging case of a multiplicative time-varying parameter.
The posterior mean estimate for $\sigma_E$ here is approximately 0.19, with 95\% of the sample between around 0.17 and 0.21.
The particle retention stays between 40\% and 60\% at most times, at the lowest dropping to 34\%.

\subsubsection*{Case 3: Estimating Both Time-Varying Stiffness and External Forcing}

In the last case, using the data set shown in Figure~\ref{Fig:Data_Oscillator}(b), we treat both $k(t)$ and $q(t)$ as unknown and aim to estimate the two time-varying parameters simultaneously.
For this experiment, we let $\sigma_C = 0.2$ and $\sigma_D = 0.5$ as before 
and test two different approaches for estimating the parameter drift: (i) estimate a single drift constant for both parameters; and (ii) estimate a different drift constant for each parameter.

\begin{figure}[t!]
  \centerline{\includegraphics[width=0.95\textwidth]{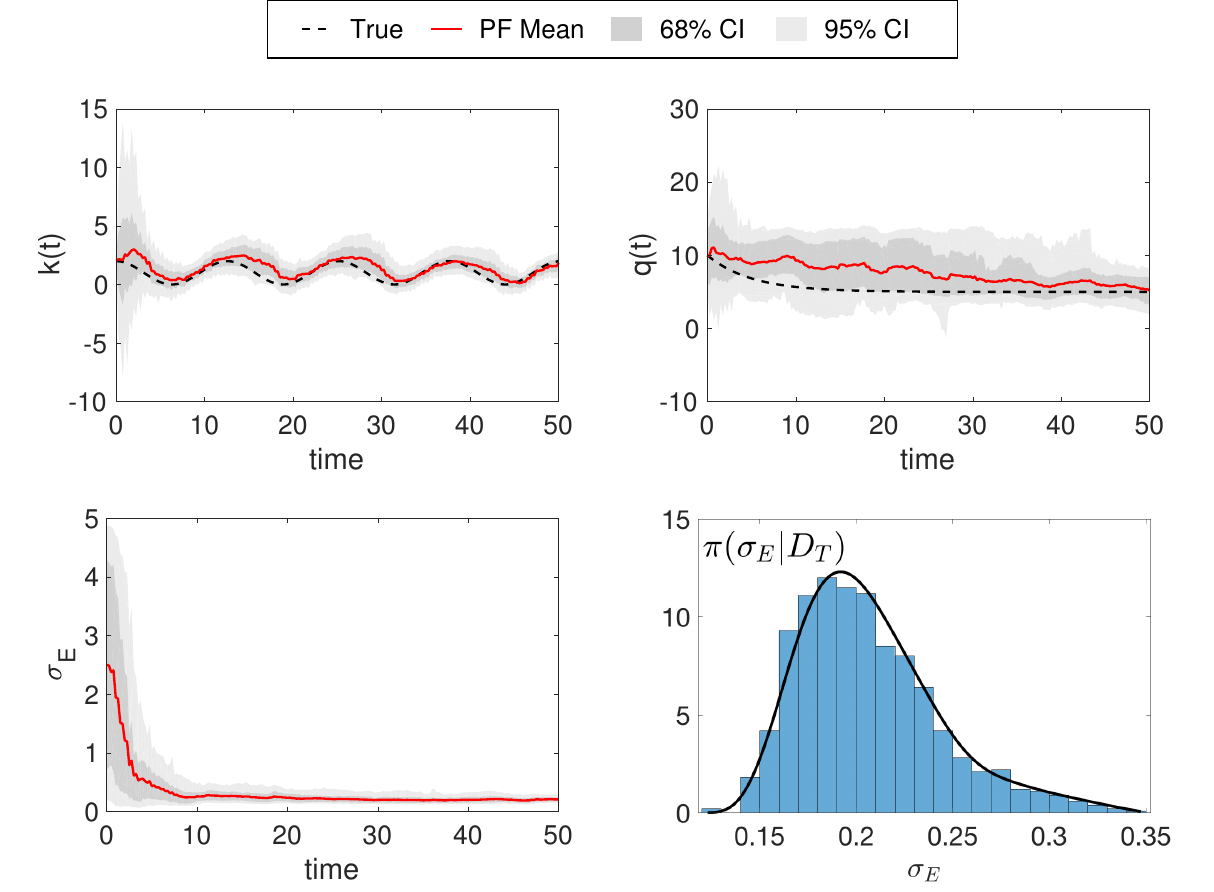} }
  \caption{Results using the PF-TVP+ algorithm to estimate both time-varying stiffness $k(t)$ and external forcing $q(t)$ in system \eqref{Eq:Forced_OscillatorP}--\eqref{Eq:Forced_OscillatorV} given the data in Figure~\ref{Fig:Data_Oscillator}(b) with one value for $\sigma_E$. Top row: Time-varying parameter estimates for $k(t)$ (left) and $q(t)$ (right). Bottom row: Time series estimate for the drift constant $\sigma_E$ (left) and histogram of the posterior sample for $\sigma_E$ at time $t=50$ (right).}
  \label{Fig:Results_Oscillator_Enhanced_Both}
\end{figure}

\begin{figure}[t!]
  \centerline{\includegraphics[width=0.95\textwidth]{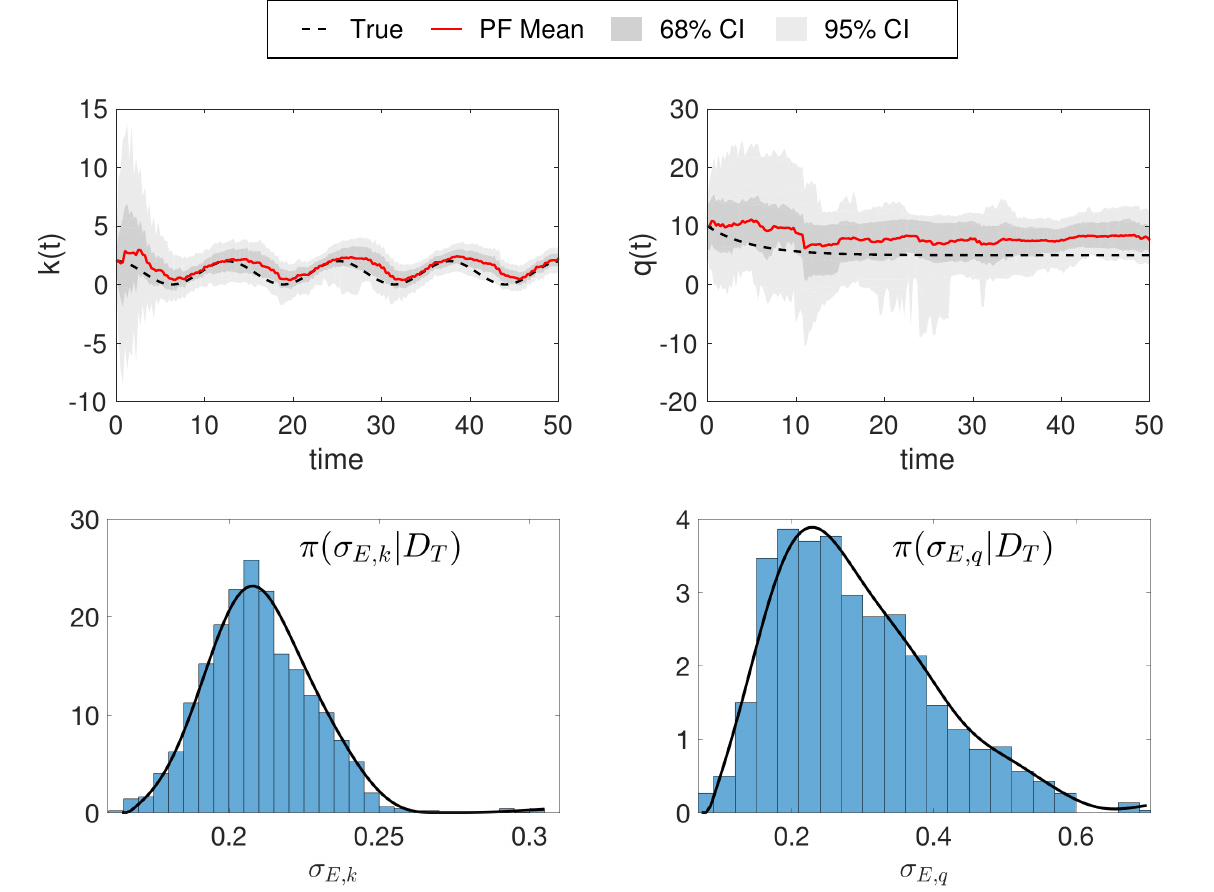}  }
  \caption{Results using the PF-TVP+ algorithm to estimate both time-varying stiffness $k(t)$ and external forcing $q(t)$ in system \eqref{Eq:Forced_OscillatorP}--\eqref{Eq:Forced_OscillatorV} given the data in Figure~\ref{Fig:Data_Oscillator}(b) with individual values for $\sigma_{E,k}$ and $\sigma_{E,q}$. Top row: Time-varying parameter estimates for $k(t)$ (left) and $q(t)$ (right). Bottom row: Histograms of the posterior samples at time $t=50$ for $\sigma_{E,k}$ (left) and $\sigma_{E,q}$ (right).}
  \label{Fig:Results_Oscillator_Enhanced_Both_2Sigmas}
\end{figure}

Figures~\ref{Fig:Results_Oscillator_Enhanced_Both} and \ref{Fig:Results_Oscillator_Enhanced_Both_2Sigmas} show the resulting PF-TVP+ estimates for the time-varying parameters, along with the corresponding estimates of the drift constants using each of the respective approaches.
In Figure~\ref{Fig:Results_Oscillator_Enhanced_Both}, we see that the algorithm is able to well track the behavior of $k(t)$ but has more difficulty retaining the exponential shape of $q(t)$; however, the algorithm does capture the decrease in $q(t)$ over time, ending right around the true function value, and maintains the true function within the 95\% credible intervals.
Note that the posterior for $\sigma_E$ has a similar mean to the resulting distribution in Figure~\ref{Fig:Results_Oscillator_Enhanced_TVStiff} when estimating $k(t)$ with fixed $q(t)$.
Here, the posterior mean estimate for $\sigma_E$ is approximately 0.21, with 95\% of the sample between values of approximately 0.15 and 0.30.

In Figure~\ref{Fig:Results_Oscillator_Enhanced_Both_2Sigmas}, we see that the estimate of $k(t)$ and its corresponding drift constant $\sigma_{E,k}$ remain similar to the previous results with a single $\sigma_E$. 
The algorithm also has similar difficulty tracking the exponential shape of $q(t)$, capturing a bit sooner the decreasing behavior but ending on a higher value than that of the true function.
The posterior mean estimate for $\sigma_{E,k}$ is approximately 0.21, with 95\% of the sample between around 0.18 and 0.24, while the posterior mean estimate for $\sigma_{E,q}$ is approximately 0.29, with 95\% of the sample between around 0.12 and 0.54.
While not shown, the model state estimates and particle retention in both of these simulations remain similar to the results shown in Figure~\ref{Fig:Results_Oscillator_Enhanced_TVStiff}.

\section{Discussion}
\label{Sec:Discussion}

This work presents a systematic particle filtering approach to estimate time-varying parameters in nonstationary inverse problems arising from deterministic dynamical systems.  
Focusing on time-varying parameter estimation in systems modeled by ODEs, we propose two particle filter algorithms utilizing random walk models to act as surrogate models for the true parameter evolution.
Numerical results in several computed examples demonstrate the capability of the algorithms in estimating time-varying parameters with different underlying functional forms and different relationships with the system states.

The performance of the PF-TVP algorithm outlined in Section~\ref{Sec:PF_TVP} relies on the a priori selection of a fixed noise covariance matrix $\mathsf{E}$ for the parameter random walk.
Results in Figure~\ref{Fig:Results_Logistic_FixedSigma} show how this choice can limit the algorithm's ability to track the underlying time-varying parameter by either over-restricting or over-inflating the parameter sample's movement so that the filter mean and credible intervals are less reliable.
To overcome this limitation, we propose an enhanced PF-TVP+ algorithm in Section~\ref{Sec:PF_TVP_Enhanced} to include online estimation of the parameter drift noise variance (more specifically, the standard deviation $\sigma_E$) along with the time-varying parameter of interest.

As shown in the computed examples in Sections~\ref{Sec:Ex_Logistic_Enhanced} and \ref{Sec:Ex_Oscillator_Enhanced}, the PF-TVP+ algorithm is able to track $\theta(t)$ and estimate $\sigma_E$ for time-varying parameters with a variety of underlying functional forms.
These include continuous functions, like the sinusoidal and linear parameters estimated for the forced logistic equation in Figures~\ref{Fig:Results_Logistic_Enhanced_Sinusoid} and \ref{Fig:Results_Logistic_Enhanced_Other}, respectively, as well as discontinuous functions, like the single and multiple step functions shown in Figure~\ref{Fig:Results_Logistic_Enhanced_Other}.
Further, the algorithm is able to estimate both additive and multiplicative time-varying parameters (see Figures~\ref{Fig:Results_Oscillator_Enhanced_Forcing}, \ref{Fig:Results_Oscillator_Enhanced_ConstStiff}, and \ref{Fig:Results_Oscillator_Enhanced_TVStiff})
and produce either a single estimate or individual estimates of $\sigma_E$ for multiple time-varying system parameters (see Figures~\ref{Fig:Results_Oscillator_Enhanced_Both} and \ref{Fig:Results_Oscillator_Enhanced_Both_2Sigmas}), as shown in the results for a forced harmonic oscillator.

When estimating unknown time-varying parameters in dynamical systems, it is important to note that while additive parameters (such as the harvesting/repopulation parameter in the logistic equation and the external forcing parameter in the harmonic oscillator) do not affect the Jacobian matrix for the system, multiplicative parameters do have an influence: uncertainty in their values causes uncertainty in the Jacobian computation and therefore affects the forward propagation of the state particles when using implicit ODE solvers such as the BDF methods. 
Future work aims to further investigate the connection between time-varying parameter uncertainty and the choice of numerical time integrator for solving the ODE system.

Another important aspect to consider is the connection between the resulting time-varying parameter estimate, the parameter drift covariance $\mathsf{E}$, and the covariance matrices $\mathsf{C}$ and $\mathsf{D}$ relating to the state evolution and observation noise processes, respectively.
These matrices directly impact the resulting particle filter estimates: $\mathsf{C}$ affects the forward model approximations of the system states, and $\mathsf{D}$ is involved in computing the likelihood and therefore affects the resampling step and particle weights.
In this work, we assume that $\mathsf{C}$ and $\mathsf{D}$ are known and fixed; however, if these matrices are unknown or poorly known, additional modifications may be needed to account for the additional uncertainty.
Future work will explore these connections and related estimation approaches.

Compared to previous works in the literature using ensemble Kalman filters (EnKFs) with random walk models for parameter evolution \cite{Campbell2020, Arnold2019}, the PF-TVP+ algorithm proposed in this work has the advantage of systematically estimating the parameter drift constant.
Future work aims to adapt this approach for use in alternative Bayesian filtering schemes, such as the augmented EnKF and its variants.
Future work also aims to extend the proposed methodology to estimate time-varying parameters in systems of partial differential equations that can be reformulated as ODEs, e.g., using a method of lines approach to discretize the spatial variables \cite{LeVeque2007}.



\section{Conclusion}
\label{Sec:Conclusions}

This work develops a systematic particle filtering approach for time-varying parameter estimation in nonstationary inverse problems arising from deterministic dynamical systems.  
The proposed method reframes the idea behind artificial parameter evolution commonly used in estimating constant parameters to instead act as a surrogate evolution model for actual time-varying parameters.
Focusing on systems modeled by ordinary differential equations, we contribute two particle filter algorithms for time-varying parameter estimation: one that relies on a fixed value for the noise variance of the parameter random walk; another that employs online estimation of the noise variance along with the time-varying parameter of interest. 
We provide several computed examples demonstrating the effectiveness of the proposed algorithms in estimating time-varying parameters with different underlying functional forms and different relationships with the system states (i.e., additive vs. multiplicative), and we highlight the considerations needed for practical implementation.
The presented framework offers a systematic computational methodology for addressing time-varying parameter estimation and has the potential for application in a wide variety of real-world estimation and prediction problems.



\section*{Acknowledgements}

This work was supported by the National Science Foundation under grant number NSF/DMS-1819203 (A. Arnold).


\section*{ORCID iD}
Andrea Arnold: \url{https://orcid.org/0000-0003-3003-882X}



\bibliography{AA_refs}{}

\end{document}